\documentclass[12pt]{article}
\usepackage{savesym}
\usepackage{cite}
\usepackage{wasysym}
\usepackage{amscd}
\usepackage{graphicx}
\usepackage{pifont}
\usepackage{float}
\usepackage{subfig}
\usepackage[all]{xy}
\usepackage{multicol}
\usepackage{amsfonts}
\usepackage{picture}
\usepackage{amssymb}
\savesymbol{iint}
\savesymbol{iiint}
\usepackage{amsmath}
\restoresymbol{TXF}{iint}
\restoresymbol{TXF}{iiint}
\usepackage{color}
\usepackage{clay}
\usepackage{hyperref}
\usepackage{slashed}
\hypersetup{colorlinks=true}
\hypersetup{linkcolor=black}
\hypersetup{citecolor=black}
\hypersetup{urlcolor=black}
\usepackage{setspace}
\usepackage{multirow}
\usepackage{wrapfig}
\usepackage{verbatim}
\usepackage{accents}
\numberwithin{equation}{section}

\begin{document}
\date{June, 2016}

\institution{Fellows}{\centerline{${}^{1}$Society of Fellows, Harvard University, Cambridge, MA, USA}}
\institution{HarvardU}{\centerline{${}^{2}$Jefferson Physical Laboratory, Harvard University, Cambridge, MA, USA}}

\title{Schur Indices, BPS Particles, and Argyres-Douglas Theories}

\authors{Clay C\'{o}rdova\worksat{\Fellows}\footnote{e-mail: {\tt cordova@physics.harvard.edu}}  and Shu-Heng Shao\worksat{\HarvardU}\footnote{e-mail: {\tt shshao@physics.harvard.edu}} }

\abstract{We conjecture a precise relationship between the Schur limit of the superconformal index of four-dimensional $\mathcal{N}=2$ field theories, which counts local operators, and the spectrum of BPS particles on the Coulomb branch.  We verify this conjecture for the special case of free field theories, $\mathcal{N}=2$ QED, and $SU(2)$ gauge theory coupled to fundamental matter.  Assuming the validity of our proposal, we compute the Schur index of all  Argyres-Douglas theories.    Our answers match expectations from the connection of Schur operators with two-dimensional chiral algebras. Based on our results we propose that the chiral algebra of the generalized Argyres-Douglas theory $(A_{k-1},A_{N-1})$ with $k$ and $N$ coprime, is the vacuum sector of the $(k,k+N)$ $W_{k}$ minimal model, and that the Schur index is the associated vacuum character. }

\maketitle

\setcounter{tocdepth}{2}
\tableofcontents

\section{Introduction}

In quantum field theory there are two distinct fundamental notions which may be formulated:  the algebra of local operators, and the Hilbert space of single-particle states.  In free field theories there are simple relations between these quantities.    A fundamental field $\phi(x)$ may act on the vacuum state to create a particle at position $x$.  Repeated action of the fundamental fields then produces multi-particle states which fill out the entire Fock space.

This simple connection between operators and particles is broken in interacting field theories, where in general the action of a local operator on the vacuum creates a complicated spectrum of states.  The situation is even more stark in the case of interacting conformal field theories which do not contain any single-particle states in their Hilbert space.

Things may be better in theories with extended supersymmetry.  In that case, there are protected local operators annihilated by some supercharges.  On the other hand, after activating moduli and spontaneously breaking conformal symmetry, there is also a well-defined particle spectrum which contains in particular BPS states.  One may then ask if there remains any connection between local operators and particles if we focus only on the subsectors of each which are protected by some supersymmetry.

In this paper we propose a sharp version of such a connection in the context of four-dimensional $\mathcal{N}=2$ field theories.  On the local operator side, we consider the so-called Schur limit of the superconformal index \cite{Kinney:2005ej,Gadde:2011ik,Gadde:2011uv}
\begin{equation}
\mathcal{I}(q)=\mathrm{Tr}\Big[(-1)^{F}q^{\Delta-\frac{1}{2}R}\Big]~. \label{shur}
\end{equation}
Here the trace is over the Hilbert space of states on $S^{3},$ or equivalently, the space of local operators.  The quantity $\Delta$ is the scaling dimension, and $R$ is the Cartan of the $SU(2)_{R}$ symmetry.  This limit of the index has been widely studied \cite{Razamat:2012uv,Razamat:2013qfa,Razamat:2014pta} due to the fact that it has enhanced supersymmtery.  It is related, in the context of class $S$ theories, to $q$-deformed Yang-Mills theory \cite{Gadde:2009kb,Gadde:2011ik,Gadde:2011uv,Kawano:2012up,Fukuda:2012jr,Mekareeya:2012tn,Tachikawa:2015iba}, and most recently has made a prominent appearance in the connection of four-dimensional $\mathcal{N}=2$ theories to two-dimensional chiral algebras \cite{Beem:2013sza,Beem:2014rza,Lemos:2014lua,Rastelli:2014jja}.

The quantity we compare to on the particle side has a more involved definition, using ingredients developed in \cite{Kontsevich:2008fj,Dimofte:2009tm,Gaiotto:2010be} and especially \cite{Cecotti:2010fi}.  The subtlety is that although any non-conformal vacuum has a well-defined set of BPS states, the exact spectrum of such particles depends on the vacuum in question and jumps at walls of marginal stability.  Since the index $\mathcal{I}(q)$ has no knowledge of the vacuum, we must therefore form a wall-crossing invariant generating function of BPS particles.   

A natural candidate emerges from wall-crossing formulas and the work of \cite{Cecotti:2010fi,Iqbal:2012xm}.  Specifically, there exists an operator $\mathcal O(q)$ valued in a quantum torus algebra.  This operator takes the schematic form
\begin{equation}
\mathcal{O}(q)\equiv \prod^{ \curvearrowright} E_{q}(X_{\gamma})~.
\end{equation}
Here the $E_{q}(X_{\gamma})$ are particular non-commutative operators and $q \sim e^{\hbar}$ is a parameter controlling the commutation relations.  There is one factor of $E_{q}(X_{\gamma})$ for each BPS state of electro-magnetic charge $\gamma$, and the product is taken in the phase order of the associated central charge.  

According to the wall-crossing formula of \cite{Kontsevich:2008fj}, crossing a wall of marginal stability the individual factors change but the operator $\mathcal{O}(q)$ is invariant.  In particular, its trace constructed via the techniques of \cite{Cecotti:2010fi} is therefore a wall-crossing invariant generating function of BPS states.  Moreover, as noticed in \cite{Cecotti:2010fi}, these traces have surprising connections to characters of two dimensional chiral algebras.  This observation will play a crucial role in the following.

With these preliminaries we may now formulate our conjectured relation between the index $\mathcal{I}(q)$ and the spectrum of BPS particles as simply
\begin{equation}
\mathcal{I}(q)=\left[\prod_{n=1}^{\infty}(1-q^{n})\right]^{2r} \mathrm{Tr}\Big[\mathcal{O}(q)\Big]~, \label{conjecture}
\end{equation}
where $r$ is the rank of the Coulomb branch.  When the theory has flavor symmetry, the index above may be further refined using flavor fugacities.\footnote{A connection of the Schur index with the traces defined in \cite{Cecotti:2011rv} has also been suggested by A. Gadde. } 

The general paradigm, that wall-crossing invariant generating functions of BPS states may be related to quantities like local operators defined naturally at the origin of moduli space originated in \cite{Cecotti:1992rm, Cecotti:2010fi, Iqbal:2012xm}.   For instance, in two-dimensional $(2,2)$ theories there is a precise connection between the BPS soliton spectrum on the moduli space and the $r$-charges of chiral operators at the origin of moduli space\cite{Cecotti:1992rm}.   A similar idea was studied in \cite{Cecotti:2010fi} where $r$-charges of chiral operators were also connected to BPS states in four-dimensions.  Our conjecture \eqref{conjecture} is inspired by those ideas, and the general philosophy of connecting moduli space physics to data at the conformal point, but differs from \cite{Iqbal:2012xm} which studies another limit of the index.\footnote{We have been told that the connection between the Schur index and the traces of the BPS monodromy operator, as a refinement of the proposal of \cite{Iqbal:2012xm}, has been recently formulated \cite{vafa}.
}
 
One interesting aspect of the conjecture \eqref{conjecture} is that it provides a connection between physics on distinct branches of moduli space.  Indeed, the generating function of BPS states is, by construction, computed on the Coulomb branch.  On the other hand, the Schur index $\mathcal{I}(q)$ can detect operators whose expectation values parameterize the Higgs branch. 

At a heuristic level, one may view our proposal \eqref{conjecture} as follows.  The prefactor multiplying the trace is the contribution of the $r$ abelian vector multiplets arising on the Coulomb branch.  Meanwhile the trace of $\mathcal{O}(q)$ selects gauge invariant combinations of BPS particles which may be produced by acting on the vacuum with BPS operators.    The conjecture \eqref{conjecture} might be amenable to analysis via supersymmetric localization, and we leave any potential derivation as an interesting problem for future work.  In the remainder of this paper we present the details of our proposal and subject it to various tests and applications.  

We begin in \S\ref{statement} with a detailed formulation of the conjecture including in particular the formalism needed to define the operator $\mathcal{O}(q)$ and its trace.  We also briefly recall the definition of the Schur index as a matrix integral in the special case of Lagrangian field theories.  We find it convenient to discuss the index $\mathcal{I}(q)$ even in the case where the four-dimensional theory in question is not conformal.  Although we lack a first principles construction of such an index, we take as a working definition in Lagrangian field theories the naive generalization of the matrix integral to this case.  The conjecture \eqref{conjecture} naturally extends to this non-conformal setting.

In \S\ref{Test} we marshall evidence for our proposal.  We evaluate independently both sides of \eqref{conjecture} in the cases of free field theories, $\mathcal{N}=2$ QED, and $SU(2)$ gauge theory coupled to fundamental matter, including all possible refinements by flavor fugacities.  In each case we find complete agreement.

Lastly, in \S\ref{AD} we consider various non-trivial consequences of the conjecture.  A particularly natural application concerns the Argyres-Douglas superconformal field theories \cite{Argyres:1995jj,Argyres:1995xn, Eguchi:1996vu, Eguchi:1996ds, Gaiotto:2010jf}.  These theories are strongly-coupled with no known Lagrangian presentation, and little is known about their operator spectrum.\footnote{The Hilbert series for the Higgs branch of some of the $(A_{k-1},A_{N-1})$ Argyres-Douglas theories is computed in \cite{DelZotto:2014kka}, which could potentially be related to the Hall-Littlewood limit of the superconformal index (see for example, \cite{Beem:2014rza}).}  They arise on the moduli space of more familiar gauge theories at special loci where electric and magnetic degrees of freedom become massless.  These field theories may also be constructed in string theory or with M5-branes \cite{Bonelli:2011aa, Xie:2012hs,Xie:2012jd,Xie:2013jc}.  On their moduli space, the Argyres-Douglas theories have particularly simple BPS spectra \cite{Shapere:1999xr,Cecotti:2010fi,Gaiotto:2010be,Cecotti:2011rv,Alim:2011ae,Alim:2011kw}, and evaluation of the BPS trace produces a natural conjecture for the Schur index.  Using this idea we produce candidate expressions for the Schur indices of all Argyres-Douglas theories.

There are a variety of consistency conditions satisfied by our proposal for the Schur indices of the Argyres-Douglas theories.  For instance, the low-order terms in the series expansion in $q$ indicate the expected operators with small scaling dimension.  In particular, we correctly recover the known global symmetries of these models.  

A sharper check on our results comes from the fact that the operators contributing to the Schur index have the structure of a two-dimensional chiral algebra \cite{Beem:2013sza,Beem:2014rza,Lemos:2014lua,Rastelli:2014jja}.   The central charges of these chiral algebras are inherited from those of the four-dimensional theory.  This in turn leads to natural proposals for the Schur index of simple Argyres-Douglas theories as the characters of certain non-unitary minimal models \cite{Lecture1, Lecture2}.   We find that such characters are exactly reproduced by our calculations.

The fact that traces of wall-crossing invariant operators $(\mathcal{O}(q))^{n}$ produce characters of two-dimensional chiral algebras featured prominently in \cite{Cecotti:2010fi}.  For instance, the trace of $\mathcal{O}(q)^{-1/2}$ gives rise to the characters of unitary coset models.  More relevant for our purposes, however, it was observed that characters of the $(2,5)$ and $(3,5)$ Virasoro minimal models arise from the traces of $\mathcal{O}(q)^{-1}$ and $\mathcal{O}(q)$ respectively in the context of the $A_{2}$ Argyres-Douglas theory, and a more general connection to non-unitary minimal models was anticipated.\footnote{See Section 9.6 of \cite{Cecotti:2010fi}.}  These relations between generating functions of BPS states and non-unitary minimal models established in \cite{Cecotti:2010fi} provide the key link in the conjecture \eqref{conjecture} that we propose.

Our work extends the observations and calculations of \cite{Cecotti:2010fi} to make contact with the specific Schur index observable at the origin of moduli space.  It would be interesting to generalize our calculations to the larger structure discovered in \cite{Cecotti:2010fi} involving traces over $\mathcal{O}(q)$ to fractional powers, and connections with the Verlinde algebra of the two-dimensional chiral algebras.   

 As a particular sampling of our results with regards to non-unitary minimal models, our conjecture implies that the $A_{2n}$ sequence of Argyres-Douglas theories, defined by Seiberg-Witten curve singularity 
\begin{equation}
y^{2}+x^{2n+1}=0~,
\end{equation}
have Schur indices given by the vacuum character of the $(2, 2n+3)$ Virasoro minimal model.  Meanwhile, for the $D_{2n+1}$ series of Argyres-Douglas models we find the vacuum character of an $\widehat{SU(2)}$  Kac-Moody algebra at level $ k_{2d}=- {4n\over 2n+1},$ and for the $D_4$ theory the vacuum character of $\widehat{SU(3)}_{-{3\over2}}$. These results agree with the conjectures of \cite{Lecture1, Lecture2}.  In $E_{6}$ and $E_{8}$ theories, we encounter the vacuum characters of the $(3,7)$ and $(3,8)$ $W_{3}$ minimal models respectively.  Based on these results we are led to propose that the chiral algebra of the generalized $(A_{k-1}, A_{N-1})$ Argyres-Douglas theory with $k$ and $N$ coprime is the vacuum sector of the non-unitary $(k,k+N)$ $W_{k}$ minimal model, and that the Schur index is the associated vacuum character.

Finally, another significant check on our results for the Schur indices of Argyres-Douglas theories comes from the recent work \cite{Buican:2015ina,Buican:2015hsa} building on \cite{Buican:2014qla, Buican:2014hfa}. Using alternative arguments, the authors propose conjectures for the Schur index of the $A_{2n+1}$ and $D_{2n}$ Argyres-Douglas theories.  In all cases we have examined our answer reproduces their expressions thereby providing additional evidence for both our work and theirs.

\section{Statement of the Conjecture}
\label{statement}
In this section we explain the exact statement of our conjecture.   We begin in \S\ref{shur} with a brief review of the Schur index.  We discuss the behavior of its low-order terms as well as its definition in Lagrangian theories as a simple integral. Moreover, we propose a naive definition for this index for non-conformal Lagrangian field theories by simply extending the standard matrix integral definition.  Next, in \S\ref{trace} we describe the wall-crossing technology needed to formulate the generating function of BPS states.  We introduce the quantum torus algebra, and the operator $\mathcal{O}(q)$ which behaves nicely under wall-crossing.  Finally, in \S\ref{substatement} we formulate the notion of trace of the operator paying particular attention to the subtleties that arise for theories with flavor symmetry.  We then state our conjecture.

\subsection{Schur Index}
\label{shur}

For the formulation of the Schur index we follow standard references \cite{Kinney:2005ej,Gadde:2011ik,Gadde:2011uv}.  In a general $\mathcal{N}=2$ conformal field theory with flavor symmetry of rank $n_f$ we may define the Schur index abstractly as a trace over the Hilbert space of states on $S^{3},$ or equivalently by the state operator correspondence, the space of local operators.  Refined by all possible flavor symmetries this takes the form
\begin{equation}
\mathcal{I}(q,z_{1}, \cdots, z_{n_f})=\mathrm{Tr}\Big[\,(-1)^{F}\,q^{\Delta-\frac{1}{2}R} \,\prod_{i=1}^{n_f} z_{i}^{f_{i}}\,\Big]~, \label{shur1}
\end{equation}
where $f_{i}$ indicate Cartans of the flavor symmetry group, $\Delta$ is the scaling dimension, and $R$ is the Cartan of the $SU(2)_{R}$ $R$-symmetry normalized to take integral values.

The index $\mathcal{I}(q)$ is protected under continuous deformations of the theory and hence is frequently computable.  A price that we pay for this simplicity is that the index does not count operators absolutely and non-trivial cancellations may occur.  The first few terms in the series expansion however, are protected against cancellations.  Specifically, from the character decompositions of \cite{Dolan:2002zh,Gadde:2011uv}  one may deduce that if the theory in question does not contain any free sector and has rank $n_f$ flavor symmetry, then the first few terms in the Schur index (at vanishing flavor fugacity) take the form
\begin{equation}
\mathcal{I}(q)=1+(n_f)q+(c_{3})q^{3/2}+(n_f+1+c_{4}-d)q^{2}+\cdots ~.\label{shurexp}
\end{equation}
This answer has several interesting features:

\begin{itemize}
\item The coefficient of $q^{0}$ counts the unique unit operator in the theory.
\item The flavor symmetry may be determined unambiguously from $\mathcal{I}(q)$.  If the index is further refined by flavor fugacities as in \eqref{shur1} then the linear term is refined to the character of the adjoint representation of the flavor group.
\item The coefficients $c_{m}$ are non-negative integers.  They count Higgs branch type operators whose primary is a Lorentz scalar in the $(m+1)$-dimensional representation of the $SU(2)_{R}$ symmetry, and with scaling dimension $m$. 
\item The coefficient of $q^{2}$ includes a contribution the flavor currents, and from the unique energy-momentum tensor multiplet of the theory (the offset by 1).  The quantity $d$ is a non-negative integer which counts the number of multiplets whose superconformal primary is a Lorentz scalar, in an $SU(2)_{R}$ triplet, with scaling dimension $\Delta=3,$ and with $U(1)_{r}$ charge $\pm 2$.\footnote{Our convention for the $U(1)_{r}$ charge is such that the supercharges have $U(1)_{r}$ equal to $\pm1.$}  
\end{itemize}
We will use this behavior of the low-order terms in the index as a simple consistency condition on our conjecture.\footnote{As a point of caution, we stress that in theories with free sectors there are additional multiplets containing for instance the free fields or higher spin currents which modify the expression \eqref{shurexp}. }

In the case of Lagrangain field theories the Schur index may be computed in the limit of zero coupling constants by a simple matrix integral.  The ingredients in this calculation are the so-called single-letter partition functions for vector and half hypermultiplets.  They are defined by
\begin{equation}
f^{V}(q)=-\frac{2q}{1-q}~, \hspace{.5in}f^{\frac{1}{2}H}=\frac{q^{1/2}}{1-q}~.
\end{equation}
We also have need of the plethystic exponential defined for a function of several variables by the operation
\begin{equation}
P.E.[f(q,z)] \equiv \exp \Bigg[\sum_{n=1}^{\infty}\frac{1}{n}f(q^{n},z^{n})\Bigg]~.
\end{equation}
Then, for a theory with gauge group $G$ and hypermultiplet matter in a representation $R$ of $G$, the (flavor refined) Schur index takes the form
\begin{equation}
\mathcal{I}(q,z)=\int [dU] \,P.E. \Big[f^{V}(q)\chi_{G}(u)+f^{\frac{1}{2}H}(q)\chi_{R}(u)\chi_{F}(z)\Big]~,\label{matrixint}
\end{equation}
where in the above, $[dU]$ denotes the Harr measure\footnote{We normalize $[dU]$ such that it has total integral one.} on the maximal torus of $G,$ and $\chi_{G},$  $\chi_{R},$ and $\chi_{F}$ are respectively the characters of the adjoint representation of $G$, the matter representation $R,$ and the representation $F$  of the flavor group.  

As described above, the Schur index $\mathcal{I}(q)$ is defined only for conformal field theories.  Nevertheless, the integral \eqref{matrixint} may clearly be defined for any $\mathcal{N}=2$ theory with a Lagrangian definition.  We propose this matrix integral as a working definition of the ``Schur index'' for the case of non-conformal $\mathcal{N}=2$ theories.  This definition is unsatisfactory since it is not clear how to extend it to general field theories, and even more worrying, it is not clear why this definition will give a protected observable.  However we will see that, at least in cases we have computed, the non-conformal extension of $\mathcal{I}(q)$ is indeed a robust observable and in particular is also captured by the generating function of BPS states as in \eqref{conjecture}.  Developing a more satisfactory theory of this non-conformal index is an important problem for future research.

\subsection{Kontsevich-Soibelman Operator $\mathcal{O}(q)$}
\label{trace}
We now turn to the particle side of our proposal.  The data entering our construction resides on the Coulomb branch of the $\mathcal{N}=2$ theory.  In this phase the physics is infrared free and governed by a $U(1)^{r}$ gauge theory.  The massive spectra may be described in terms of particles carrying various electric, magnetic, and flavor charges.  These charges reside in an integral lattice $\Gamma$ equipped with an integer-valued antisymmetric Dirac pairing that we denote by $\langle \cdot, \cdot \rangle$.

A crucial role is played by the central charge $\mathcal{Z}$.  For each fixed vacuum modulus, $\mathcal{Z}$ is a complex-valued linear function on $\Gamma$.  Its importance lies in the fact that the mass of any single-particle state of charge $\gamma \in \Gamma$ obeys 
\begin{equation}
M \geq |\mathcal{Z}(\gamma)|~.
\end{equation}
Particles whose masses saturate the above bound are BPS.  They are annihilated by some of the supersymmetries.  As a result, they may be frequently computed even at strong coupling using a variety of distinct methods.  

For each fixed charge $\gamma \in \Gamma$ the spectrum of one-particle BPS states is captured by an index, the protected spin character \cite{Gaiotto:2010be}.   The particles are in representations of the $SU(2)_{J}$ little group and the $SU(2)_{R}$ $R$-symmetry group.  Upon factoring out the center-of-mass hypermultiplet, the one-particle Hilbert space of charge $\gamma$ is 
\begin{equation}
H_{\gamma}=\Big[ (\mathbf{2},\mathbf{1})\oplus  (\mathbf{1},\mathbf{2})\Big] \otimes H_{int}(\gamma)~,
\end{equation}
where $H_{int}(\gamma)$ is some representation of $SU(2)_{J}\times SU(2)_{R}$ which encodes the internal degrees of freedom of the particles.   The protected spin-character is then the trace 
\begin{equation}
\Omega(\gamma,y)=\mathrm{Tr}_{H_{int}(\gamma)}\Big[y^{J}(-y^{R})\Big]=\sum_{n\in \mathbb{Z}}\Omega_{n}(\gamma)y^{n}~, \label{omegadef}
\end{equation}
where $J$ is the Cartan of the $SU(2)_{J}$ little group, and $R$ is the Cartan of the $SU(2)_{R}$ symmetry.  The quantities $\Omega_{n}(\gamma)$ are thus integers which encode the spin content of BPS particles of charge $\gamma$.

The spectrum of BPS particles (and hence the collection of integers $\Omega_{n}(\gamma)$) is stable under small deformations in parameters.  However, as moduli are varied the phases of central charges for distinct $\gamma$'s may align.  These loci are the walls of marginal stability.  When crossing such a wall, the BPS spectrum may jump.  The changes in the BPS degeneracies are controlled by wall-crossing formulas \cite{Kontsevich:2008fj,Dimofte:2009tm} which we now review.

To begin, for each $\gamma \in \Gamma,$ we introduce a formal variable $X_{\gamma}$.  These variables obey  a  \textit{quantum torus algebra},
\begin{equation}
X_{\gamma}X_{\gamma'}=q^{\frac{\langle\gamma,\gamma '\rangle }{2}}X_{\gamma+\gamma'}=q^{\langle\gamma,\gamma' \rangle }X_{\gamma'}X_{\gamma}~. \label{qtorus}
\end{equation}
In this equation $q$ is a formal variable controlling the non-commutativity of the algebra.  In our final proposal relating BPS particles and indices, we will see that $q$ is reinterpreted as the Schur index fugacity parameter.  We also define a function, the \textit{$q$-exponential}, as
\begin{equation}
E_{q}(z)=\prod_{i=0}^{\infty}(1+q^{i+\frac{1}{2}}z)^{-1}=\sum_{n=0}^{\infty} \frac{(-q^{\frac{1}{2}}z)^{n}}{(q)_n}~. \label{qexp}
\end{equation}
where we use the standard definitions of the $q$-Pochhammer symbols
\begin{equation}
(q)_{n}\equiv
\begin{cases}
 1~~~~~~~~~~~~~~~~~~~~~\text{if}~~n=0\,,\\
 \prod_{k=1}^{n}(1-q^{k})~~~~~\text{if}~~n>1\,.
 \end{cases}
\end{equation}

The basic fact that makes wall-crossing formulas possible is that, when evaluated on elements of the quantum torus algebra, products of the functions $E_{q}(z)$ obey remarkable identities.  The simplest of these is 
\begin{equation}
E_{q}(X_{\gamma_{1}})E_{q}(X_{\gamma_{2}})=E_{q}(X_{\gamma_{2}})E_{q}(X_{\gamma_{1}+\gamma_{2}})E_{q}(X_{\gamma_{2}})~,
\end{equation}
for charges $\gamma_{i}$ with $\langle \gamma_{1}, \gamma_{2}\rangle =1$.  This equality encodes the most simple wall-crossing process where a single hypermultiplet disappears across a wall.

Returning to the general story, we may now phrase the consequence of the wall-crossing formula which we require.  For each charge $\gamma\in \Gamma$ we build the following element of the quantum torus algebra using the $q$-exponential \eqref{qexp} and the protected spin character \eqref{omegadef}
\begin{equation}
U_{\gamma}=\prod_{n\in \mathbb{Z}}E_{q}((-1)^{n}q^{n/2}X_{\gamma})^{(-1)^{n}\Omega_{n}(\gamma)}~.
\end{equation}
Note that in simple chambers where there are only hypermultiplets, all of the $U_{\gamma}$ reduce to $q$-exponentials.

We then form an element $\mathcal{O}(q)$ in the quantum torus algebra by taking a product over all the $U_{\gamma}$ 
\begin{equation}
\mathcal{O}(q)\equiv \prod_{\gamma \in \Gamma}^{ \curvearrowright}U_{\gamma}~. \label{odef}
\end{equation}
Since the quantum torus algebra is non-commutative we must prescribe a specific order to the above product.  This is achieved with the central charge.  Specifically, we pick an arbitrary phase $\theta$ in the central charge plane.  We then define $\mathcal{O}(q)$ to be the product over all the $U_{\gamma}$ taken in order of the phase of $Z(\gamma)$ with operators of smallest phase on the left.  Defined in this way the resulting operator $\mathcal{O}(q)$ depends upon the initial phase $\theta$.  We will ultimately see that this ambiguity drops out of our index prescription.

We refer to the element $\mathcal{O}(q)$ as the \textit{Kontsevich-Soibelman (KS) operator}.  Because of CPT invariance, the product over the first sector of phase $\pi$ in the central charge plane may be interpreted as arising from particles, while the product over the second sector of total phase $\pi$ may be interpreted as arising from antiparticles.  The KS operator therefore samples the entire BPS particle spectrum of the theory. The content of the wall-crossing formula is that $\mathcal{O}(q)$ is in fact independent of the Coulomb branch vacuum.  Therefore, as observed by \cite{Cecotti:1992rm,Cecotti:2010fi,Iqbal:2012xm}, any quantity constructed out of  $\mathcal{O}(q)$ has a chance to reproduce aspects of the quantum field theory defined at the origin of moduli space.

Let us also clarify the relation of our operator $\mathcal{O}(q)$ to the BPS monodromy $M(q)$ appearing in \cite{Cecotti:2010fi}.  Up to insignificant details in conventions, we have the relation
\begin{equation}
\mathcal{O}(q)=M(q)^{-1}~.
\end{equation}
Both expressions are wall-crossing invariant and may be used to define invariant quantities at the origin of moduli space following the general philosophy of \cite{Cecotti:1992rm,Cecotti:2010fi,Iqbal:2012xm}.    In \cite{Cecotti:2010fi} traces of various powers, positive, negative, and even fractional, of $\mathcal{O}(q)$ were considered and connected to characters of two-dimensional chiral algebras.  Our conjecture relates traces of  $\mathcal{O}(q)$ to the Schur index. It would be interesting to find similar direct interpretations of traces of these various powers in terms of local operator counting at the origin of moduli space.
\subsection{Trace of $\mathcal{O}(q)$ and the Schur Index}
\label{substatement}

We are at last in a position to precisely define the BPS generating function that we claim is related to the Schur index.  We use the notions of trace defined in \cite{Cecotti:2010fi}.

We begin in the simplest case where there is no flavor symmetry and subsequently relax this assumption.   We define a trace operation on the quantum torus algebra by specifying its action on generators and extending linearly
\begin{equation}
\mathrm{Tr}[X_{\gamma}]=\begin{cases} 1  & \gamma=0 ~,  \\
0 & \mathrm{else}~.
\end{cases}
\end{equation}
Note that with this definition the trace is cyclic (for instance $\mathrm{Tr}[X_{\gamma}X_{\gamma'}]=\mathrm{Tr}[X_{\gamma'}X_{\gamma}]$).  This cyclic property enables us to form an unambiguous notion of the trace of the operator $\mathcal{O}(q)$.  Indeed recall from the definition \eqref{odef}, that $\mathcal{O}(q)$ depends on an initial phase $\theta$.  As $\theta$ is varied, the factors $U_{\gamma}$ reorder cyclicly so that the trace is unmodified.

With this understanding, we may now state our conjectured relation between the Schur index and the trace of $\mathcal{O}(q)$ as
\begin{equation}
\mathcal{I}(q)=(q)_{\infty}^{2r}~\mathrm{Tr}\Big[\mathcal{O}(q)\Big]~, \label{noflavorconj}
\end{equation}
where as above $r$ is the rank of the Coulomb branch.

When the $\mathcal{N}=2$ theory has flavor symmetry the notion of trace and the statement of the conjecture must be refined.  Flavor charges enter into the discussion as elements of the charge lattice $\Gamma$ with trivial Dirac pairings.  Thus, $\gamma$ is a flavor charge if and only if for all $\gamma' \in \Gamma$ we have
\begin{equation}
\langle \gamma, \gamma' \rangle =0~.
\end{equation}
It then follows from the definition of the quantum torus algebra \eqref{qtorus} that if $\gamma$ is a flavor charge the associated element $X_{\gamma}$ is central i.e.
\begin{equation}
X_{\gamma} X_{\gamma'}=X_{\gamma'} X_{\gamma}~, \hspace{.5in}\forall \gamma' \in \Gamma~. 
\end{equation}

To extend the notion of trace to this case, we allow the traces of the flavor charges to be general non-vanishing numbers, and we define the trace of any element in terms of these choices.  Thus, pick an integral basis $\gamma_{f_{i}}\in \Gamma$ for the flavor charges.  For a general element $X_{\gamma}$ we define
\begin{equation}
\mathrm{Tr}[X_{\gamma}]=\begin{cases}  \prod_{i}\mathrm{Tr}[X_{\gamma_{f_{i}}}]^{f_{i}(\gamma)}  & \langle\gamma, \gamma'\rangle=0 ~ \forall ~ \gamma' \in \Gamma~, \\
0 & \mathrm{else}~,
\end{cases}
\end{equation}
where $f_{i}(\gamma)$ are the flavor charges of $\gamma$.  So defined, the trace of any element of the quantum torus algebra, (in particular $\mathcal{O}(q)$) is a function of $q$ and of the $n_f$ variables $\mathrm{Tr}[X_{\gamma_{f_{i}}}]$ where $i=1, \cdots, n_f$ runs over a basis of the flavor charges.

We may now again connect the trace of $\mathcal{O}(q)$ to the flavor refined Schur index.  We conjecture that 
\begin{equation}
\mathcal{I}(q, z_{1}, \cdots, z_{n})=(q)_{\infty}^{2r}~\mathrm{Tr}\Big[\mathcal{O}(q)\Big]\left(\mathrm{Tr}[X_{\gamma_{f_{1}}}],\cdots, \mathrm{Tr}[X_{\gamma_{f_{i}}}] \right)~.\label{conjflav}
\end{equation}
To fully specify this proposal, the flavor fugacities $z_{i}$ in the index must be related to traces of the flavor generators.  Thus, for each $i$ we must specify functions $h_{i}$ as 
\begin{equation}
\mathrm{Tr}[X_{\gamma_{f_{i}}}]=h_{i}(z_{1}, \cdots, z_{n_f})~.
\end{equation}
The functions $h_{i}(z)$ are model (and basis) dependent.  In practice they may be fixed by examining the low-order terms in the index and matching, for instance, the linear coefficient in $q$ to the character of the adjoint representation as explained in  \eqref{shurexp}.\footnote{In examples it is also the case that the functions $h_{i}$ do not in general evaluate to one at $z_{i}=1$. Thus the $h_{i}$ must be determined even if one wishes to compute the unrefined index.}  

Before concluding we should also note a significant technical point that requires further development.  The trace operations defined here can be readily evaluated on finite sums of elements of the quantum torus algebra.  However the operator $\mathcal{O}(q)$ whose trace we desire is an infinite sum.  In practice, the only way that we know to evaluate the trace is to expand the infinite sums and to commute the order of trace and summation.  While in the simplest examples that we describe in \S\ref{Test} and \S\ref{AD} this gives a well-behaved answer, in more complicated examples the resulting sums do not appear to converge absolutely.  This obstructs us from testing our conjecture in these examples. For instance $SU(2)$ $N_{f}=4$ Yang-Mills is of this type.  To understand these BPS traces more directly in such examples is necessary to make \eqref{conjflav} into a universally calculable tool.

\section{Tests of the Proposal}
\label{Test}

In this section we study simple examples of the conjectured relations \eqref{noflavorconj} and \eqref{conjflav} between the Schur index and the trace of $\mathcal{O}(q)$.  In all theories where we have independently computed both quantities we find perfect agreement.  This is true even in examples which are not conformal, where we use the matrix integral \eqref{matrixint} to extend the definition of the Schur index to this case. 
 
We begin in \S\ref{free} by considering the simplest models of free field theories.  In these cases, the quantum torus algebra is trivial and the relationship can be proven using simple identities obeyed by the $q$-exponential.  Next in \S\ref{qed}, we consider the infrared free theory of $\mathcal{N}=2$ QED.  This is the simplest example with non-trivial torus algebra and again the conjecture can be proven using functional identities.  Finally in \S\ref{su2}, we study examples of $SU(2)$ gauge theory coupled to fundamental matter.  We consider the cases $N_{f}=0,1,2,3$ and verify the conjecture including the dependence on flavor fugacities.\footnote{As explained in \S\ref{substatement}, the case $N_{f}=4$ cannot be evaluated using our current understanding since the trace of $\mathcal{O}(q)$ does not absolutely converge.}
\subsection{Free Field Theories}
\label{free}

\subsubsection{$U(1)$ Vector Multiplet}

Let us start by considering the free theory of a $U(1)$ vector multiplet. 
 This theory has no flavor symmetry. There are no massive BPS states so the KS operator is trivial. The Schur index is given by
\begin{align}
\mathcal{I}^V(q)  = P.E.\left[ f^V(q) \right] =  \exp \left[ \sum_{n=1}^\infty {1\over n }  {-2q^n \over 1-q^n}\right] \,,
\end{align}
where $f^V(q) = {-2q\over 1-q}$ is the single-letter partition function for a vector multiplet  and $P.E.$ is the plethystic exponential $P.E.\left[ f(q,z) \right]  =  \exp\left[ \sum_{n=1}^\infty {1\over n} f( q^n ,z^n)\right]$. We can rewrite the exponent as
\begin{align}
\sum_{n=1}^\infty {1\over n }  {-2q^n \over 1-q^n} =-2 \sum_{n=1}^\infty {q^n\over n}\sum_{k=0}^\infty q^{kn} = 2\sum_{k=0}^\infty  \log (1-q^{k+1}) =2 \log[  (q)_\infty]\,.
\end{align}
Therefore, we find agreement with our conjectured relation \eqref{noflavorconj}
\begin{align}
\mathcal{I}^V(q)  = (q)_\infty^2~.
\end{align}

\subsubsection{Free Hypermultiplet}

Having checked the  free theory of a single vector multiplet, let us move on to the case of a  free hypermultiplet. This theory has an $Sp(1)\cong SU(2)$ flavor symmetry whose fugacity will be denoted by $z$.  The charge lattice $\Gamma$ of this free theory is real, and one-dimensional corresponding to the $SU(2)$ flavor symmetry and the quantum torus algebra is commutative. The only BPS particle  is the hypermultiplet state itself and there is no wall-crossing phenomenon. We denote the lattice vector of this hypermultiplet state by $\gamma\in\Gamma$ and the corresponding generator in the quantum torus algebra by $X_\gamma$. We normalize the trace of the generator as
\begin{align}
 \text{Tr}[X_\gamma]= -z\,.
 \end{align}

The KS operator is given by
\begin{align}
\mathcal{O}(q)  =  E_q( X _{-\gamma}) E_q(X_\gamma)\,.
\end{align}
Its trace is obtained by replacing $X_\gamma$ by $-z$,
\begin{align}
\text{Tr} \left[ \mathcal{O}(q) \right] = E_q (-z^{-1}  ) E_q(-z)\,.
\end{align}
Using the identity
\begin{align}\label{dilogid}
E_q(-z):=\prod_{i=0}^\infty (1-q^{i+{1\over2}} z)^{-1}=\exp \left[ \sum_{n=1}^\infty {1\over n}{ q^{n\over2}  \over 1-q^n }  z^n\right]\,,
\end{align}
we can rewrite the trace as
\begin{align}
\text{Tr} \left[ \mathcal{O}(q) \right]= \exp \left[ \sum_{n=1}^\infty {1\over n}{ q^{n\over2}  \over 1-q^n } (z^n +z^{-n})\right]\,.
\end{align}

On the other hand, the Schur index of a free hypermultiplet is given by
\begin{align}
\mathcal{I}_{H} (q,z)=P.E.\left[ f^{{1\over 2} H} (q) (z+z^{-1})  \right]  = \exp \left[ \sum_{n=1}^\infty {1\over n}{ q^{n\over2}  \over 1-q^n } (z^n +z^{-n})\right]\,,
\end{align}
where $f^{ {1\over2} H}(q) = {q^{1\over2} \over 1-q}$ is the single-letter partition function for a half-hypermultiplet. Indeed, we see that the our proposal 
\begin{align}
\mathcal{I}^H (q,z)= \text{Tr}\left[ \mathcal{O}(q)\right]\,,
\end{align}
is satisfied. Note that there is no Coulomb branch in the free hypermultiplet theory, $r=0$, so the  prefactor  $(q)_\infty^{2r}$ in \eqref{conjflav} is trivial.

\subsection{QED}
\label{qed}
Consider the non-conformal theory of a $U(1)$ vector multiplet coupled to a hypermultiplet with one unit of $U(1)$ charge. The would-be flavor symmetry of the hypermultiplet is gauged so there is no flavor symmetry left. The Coulomb branch of the theory is complex one-dimensional and the charge lattice $\Gamma$ has real dimension two.  The only BPS particle in this theory is the hypermultiplet state. We will denote the  charge lattice vector for the hypermultiplet by $\gamma$ and the corresponding generator by $X_\gamma$. Since there is a nontrivial Dirac pairing on the charge lattice $\Gamma$, \textit{i.e.} $\gamma$ does not commute with every element in $\Gamma$, the trace of $X_\gamma$ is zero, $\text{Tr}[X_\gamma]=0$.

 The KS operator is given by
\begin{align}
\mathcal{O}(q) = E_q(X_{-\gamma} ) E_q(X_\gamma)\,.
\end{align}
The trace projects out the constant term in $X_\gamma$,
\begin{align}
\text{Tr} [\mathcal{O}(q) ]  =\sum_{k,\ell=0}^\infty\text{Tr}\left[ {(-1)^{k+\ell}q^{ {k+\ell\over2}} \over (q)_k (q)_\ell  }  X_{-k\gamma} X_{\ell \gamma}\right]
= \sum_{\ell=0}^\infty { q^\ell\over (q)_\ell^2} \,.
\end{align}

Meanwhile, the index for this non-conformal theory computed at zero coupling is
\begin{align}
\begin{split}
\mathcal{I}_{\text{QED}} (q)  &=   \int_0^{2\pi } {d\theta\over 2\pi}\,  P.E. \left[ f^V(q)  + f^{{1\over2}H} (q) (e^{i\theta} + e^{-i\theta}) \right]\,\\
& =(q)_\infty^2 \int_0^{2\pi } {d\theta\over 2\pi} \exp\left[ \sum_{n=1}^\infty {1\over n} {q^{n\over2} \over 1-q^n } (e^{i\theta} + e^{-i\theta})\right]\,.
\end{split}
\end{align}
We can evaluate the integral in $\theta$ more explicitly as follows.  Using \eqref{dilogid}, we have
\begin{align}
\begin{split}
\mathcal{I}_{\text{QED}} (q) &= (q)_\infty^2  \int_0^{2\pi } {d\theta\over 2\pi}\,
E_q(-e^{-i\theta}) E_q(-e^{i\theta})
=(q)_\infty^2 \sum_{k,\ell=0}^\infty { q^{ {k+\ell\over2}} \over (q)_k(q)_\ell} \int_0^{2\pi } {d\theta\over 2\pi}\,e^{i(-k+\ell)\theta} 
\\&= (q)_\infty^2\sum_{\ell=0}^\infty  {  q^\ell \over (q)_\ell^2} \,. 
\end{split}
\end{align}

Hence we have verified that
\begin{align}
\mathcal{I}_{\text{QED}} (q) = (q)_\infty^2 ~\text{Tr} [\mathcal{O}(q)]~.
\end{align}

\subsection{$SU(2)$ with Matter}
\label{su2}

The Schur index evaluated at the zero coupling point for the $\mathcal{N}=2$ $SU(2)$ theory with $N_f$ fundamental hypermultiplets can be uniformly written as
\begin{align}
\mathcal{I}_{SU(2),N_f}  (q) ={1\over \pi}\int_0^{2\pi}  d\theta\, \sin^2\theta\,P.E. \left[ f^V(q) \chi_{\bf 3}^{SU(2)}(\theta)+  N_f f^{{1\over 2}H} (q)  \chi_{\bf 2}^{SU(2)}(\theta)\right]\,,
\end{align}
where $\chi_{\bf 2}^{SU(2)} (\theta) = e^{i\theta}+e^{-i\theta}$ and $\chi_{\bf 3}^{SU(2)}(\theta) = e^{2i\theta}+ e^{-2i\theta}+1$ are the characters for $SU(2)$ in the fundamental and adjoint representation respectively, and ${1\over \pi }\sin^2\theta d\theta$ is the normalized Haar measure.  

In the case $N_f>0$, the theory has an $SO(2N_f)$ flavor symmetry and  we will  further weight the index by the corresponding fugacities. We will consider the case where the bare mass of the hypermultiplet is zero and the central charges of the BPS particles therefore satisfy certain linear relations. 

Throughout, we evaluate the trace of $\mathcal{O}(q)$ in the strong-coupling chamber where there are only a finite number of hypermultiplets.  The charges of these states are conveniently encoded by the BPS quiver \cite{Cecotti:2010fi, Cecotti:2011rv, Alim:2011kw, Alim:2011ae}.  Each node of the quiver represents a hypermultiplet and the Dirac pairing $\langle \gamma_{i}, \gamma_{j}\rangle $  is read off from the number of arrows from node $i$ to node $j$.\footnote{Note that the quiver encodes only the particles, there are also an equal number of antiparticles with opposite charges.  The fact that the nodes of the quiver are the only stable states is special to this particular chamber of moduli space.  In other chambers, non-trivial bound states of the node particles exist.  See e.g. \cite{Cordova:2014oxa, Hori:2014tda} for a systematic calculation for both the gauge theory and Argyres-Douglas examples. }

\subsubsection{Pure $SU(2)$ }

\begin{figure}[h]
\begin{center}
\includegraphics[width=.2\textwidth]{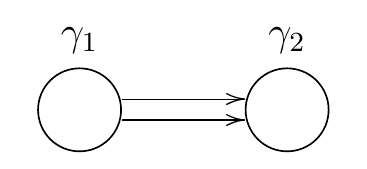}
\end{center}
\caption{ The BPS quiver for the $\mathcal{N}=2$ pure $SU(2)$ gauge theory.}\label{fig:Nf0}
\end{figure}

The $\mathcal{N}=2$ $SU(2)$ gauge theory with no matter is our first example where there is a nontrivial wall-crossing phenomenon. The BPS quiver is shown in Figure \ref{fig:Nf0}. There are two chambers in the Coulomb branch. The weak-coupling chamber defined by $\text{arg}\, \mathcal{Z}(\gamma_1)>\text{arg} \,\mathcal{Z}(\gamma_2)$ has infinitely many BPS particles, whereas the strong-coupling chamber 
\begin{align}
\text{arg}\, \mathcal{Z}(\gamma_2)>\text{arg} \,\mathcal{Z}(\gamma_1)~,
\end{align} 
has only two particles,  the monopole $\gamma_1$ and the dyon $\gamma_2$. Their Dirac pairing is $\langle \gamma_1,\gamma_2\rangle=2$. There is no flavor symmetry in this theory.

The KS operator written in the strong coupling chamber is
\begin{align}
\mathcal{O}(q) = E_q(X_{-\gamma_1})E_q(X_{-\gamma_2})  E_q(X_{\gamma_1}) E_q(X_{\gamma_2})\,.
\end{align}
The trace of the KS operator can be computed as follows
\begin{align}
\begin{split}
\text{Tr} [\mathcal{O}(q)]  &=\sum_{k_1,k_2,\ell_1,\ell_2=0}^\infty {(-1)^{k_1+k_2+\ell_1+\ell_2} \,q^{{1\over2 }(k_1+k_2+\ell_1+\ell_2) }  \over   (q)_{k_1} (q)_{k_2} (q)_{\ell_1}(q)_{\ell_2} }
\text{Tr} \left[ 
X_{-\gamma_1}^{k_1} X_{-\gamma_2}^{k_2} X_{\gamma_1}^{\ell_1} X_{\gamma_2}^{\ell_2}
\right]\\
&=\sum_{k_1,k_2,\ell_1,\ell_2=0}^\infty {(-1)^{k_1+k_2+\ell_1+\ell_2} \,q^{{1\over2}(k_1+k_2+\ell_1+\ell_2 ) }  \over   (q)_{k_1} (q)_{k_2} (q)_{\ell_1}(q)_{\ell_2} }\, q^{ 2\ell_1k_2}\,
\text{Tr} \left[ 
X_{\gamma_1}^{-k_1+\ell_1} X_{\gamma_2}^{-k_2+\ell_2}
\right]\,,
\end{split}
\end{align}
where we have used the commutation relation in the quantum torus algebra \eqref{qtorus}  $X_{\gamma_1} X_{\gamma_2}  = q^{\langle \gamma_1,\gamma_2\rangle} X_{\gamma_2} X_{\gamma_1}$. Finally, by noting that Tr$[X_{\gamma_1}]=\text{Tr}[X_{\gamma_2}]=0$, we have
\begin{align}
\text{Tr} [\mathcal{O}(q)]  = \sum_{\ell_1,\ell_2=0}^\infty { q^{\ell_1+\ell_2 +2\ell_1\ell_2 }  \over   [(q)_{\ell_1}(q)_{\ell_2}]^2 }\,.
\end{align}
We therefore obtain a $q$-expansion for
\begin{align}
 (q)_\infty^2 \text{Tr} [\mathcal{O}(q)]  =  1+q^2 +q^6 +q^{12}+q^{20}+\cdots\,.
 \end{align}

On the other hand, the Schur index is given by
\begin{align}
\begin{split}
\mathcal{I}_{SU(2),N_f=0}  (q)& =  \,{1\over \pi}\int_0^{2\pi}  d\theta\, \sin^2\theta\, P.E. \left[ f^V(q)(e^{2i\theta}+e^{-2i\theta}+1) \right]\\
&=1+q^2 +q^6 +q^{12}+q^{20}+\cdots\,,
\end{split}
\end{align}
which indeed agrees with our proposal
\begin{align}
\mathcal{I}_{SU(2),N_f=0}  (q)   = (q)_\infty^2 \text{Tr} [\mathcal{O}(q)] \,.
\end{align}

\subsubsection{$SU(2)$ with $N_f=1$}

\begin{figure}[h]
\begin{center}
\includegraphics[width=.2\textwidth]{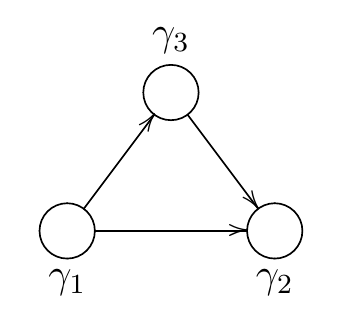}
\end{center}
\caption{ The BPS quiver for the $\mathcal{N}=2$ $SU(2)$ gauge theory with one hypermultiplet in the fundamental representation.}\label{fig:Nf1}
\end{figure}

The $\mathcal{N}=2$ $SU(2)$ gauge theory with one hypermultiplet in the fundamental representation has an $SO(2)$ flavor symmetry, whose fugacity will be denoted by $z$. The BPS quiver is shown in Figure \ref{fig:Nf1}.  There is a strong coupling chamber with the following BPS phase order 
\begin{align}
\text{arg}\, \mathcal{Z}(\gamma_2) >\text{arg}\,\mathcal{Z}(\gamma_3)>\text{arg}\, \mathcal{Z}(\gamma_1)\,,
\end{align}
 where the only BPS particles are the nodes $\gamma_1$, $\gamma_2$, $\gamma_3$. The lattice vector $\gamma_f$ corresponding to the  flavor symmetry is
 \begin{align}
 \gamma_f = \gamma_1+\gamma_2-\gamma_3\,.
 \end{align}
 By definition, $\langle \gamma_f , \gamma\rangle=0$ for all $\gamma \in \Gamma$. We normalize the trace of the flavor generator as
\begin{align}
\text{Tr}[X_{\gamma_f}] = z^2\,.
\end{align}

The KS operator is
\begin{align}
\mathcal{O}(q) = E_q (X_{-\gamma_1})E_q (X_{-\gamma_3})E_q (X_{-\gamma_2}) E_q (X_{\gamma_1})E_q (X_{\gamma_3})E_q (X_{\gamma_2})\,.
\end{align}
By expanding out $E_q(x)$, the trace of the KS operator becomes
\begin{align}\label{Nf=1tr1}
\begin{split}
\text{Tr}[\mathcal{O}(q)] &=\sum_{ \substack{\ell_1,\ell_2,\ell_3 ,\\k_1,k_2,k_3=0}}^\infty
{   (-1)^{\sum_{i=1}^3(k_i+\ell_i) } q^{{1\over2} \sum_{i=1}^3(k_i+\ell_i) } \over \prod_{i=1}^3(q)_{k_i} (q)_{\ell_i}  }
\text{Tr} \left[
 X_{-\gamma_1}^{k_1} X_{-\gamma_3}^{k_3} X_{-\gamma_2}^{k_2} 
  X_{\gamma_1}^{\ell_1} X_{\gamma_3}^{\ell_3} X_{\gamma_2}^{\ell_2} 
\right]\,.
\end{split}
\end{align}
We can compute the trace of six generators with the help of the quantum torus algebra \eqref{qtorus},
\begin{align}\label{Nf=1tr2}
\begin{split}
 X_{-\gamma_1}^{k_1} X_{-\gamma_3}^{k_3} X_{-\gamma_2}^{k_2} 
  X_{\gamma_1}^{\ell_1} X_{\gamma_3}^{\ell_3} X_{\gamma_2}^{\ell_2} 
&=q^{(\ell_1+\ell_3)k_2}\,
 X_{-\gamma_1}^{k_1} X_{-\gamma_3}^{k_3} 
  X_{\gamma_1}^{\ell_1} X_{\gamma_3}^{\ell_3} X_{\gamma_2}^{-k_2+\ell_2} 
\\
&=q^{(\ell_1+\ell_3)k_2+\ell_1k_3}\,
  X_{\gamma_1}^{-k_1+\ell_1} X_{\gamma_3}^{-k_3+\ell_3} X_{\gamma_2}^{-k_2+\ell_2} 
\,.
\end{split}
\end{align}
Finally, using $\gamma_f=\gamma_1+\gamma_2-\gamma_3$ we can write
\begin{align}
X_{\gamma_3} ^n= X_{\gamma_1+\gamma_2-\gamma_f}  ^n= q^{-{1\over2}n^2 } X_{\gamma_1} ^n X_{\gamma_f}^{-n}\,.
\end{align}
Plugging this into \eqref{Nf=1tr1} and \eqref{Nf=1tr2} we obtain 
\begin{align}
\begin{split}
&\text{Tr}[\mathcal{O}(q)]\\
&
=\sum_{ \substack{\ell_1,\ell_2,\ell_3 ,\\k_1,k_2,k_3=0}}^\infty
{   (-1)^{\sum_{i=1}^3(k_i+\ell_i) } q^{{1\over2} \sum_{i=1}^3(k_i+\ell_i)-{1\over2}(k_3-\ell_3)^2 +(\ell_1+\ell_3)k_2+\ell_1k_3 } \over \prod_{i=1}^3(q)_{k_i} (q)_{\ell_i}  } z^{2(k_3-\ell_3)}\, \delta_{k_1+k_3,\ell_1+\ell_3}\delta_{k_2+k_3,\ell_2+\ell_3}~.\\
\end{split}
\end{align}
We therefore obtain a $q$-expansion for
\begin{align}
&(q)_\infty^2 \text{Tr}[\mathcal{O}(q) ]  = 1+  q + \left( -{1\over z^2} +2 -z^2 \right) q^2 + 
\left( -{1\over z^2} +2 -z^2\right)q^3 + \left( -{2\over z^2} +4- 2z^2\right)  q^4   \nonumber \\
&~~~~~~~~~~~~+  \left( {1\over z^4}  - {3\over z^2} +5 -3z^2+z^4\right)q^5
+\left( {1\over z^4}  -{5\over z^2} +8 -5z^2 +z^4 \right)q^6
+\cdots\,.
\end{align}

On the other hand, the Schur index computed at the zero coupling is
\begin{align}
&\mathcal{I}_{SU(2),N_f=1} (q,z) = {1\over \pi} \int^{2\pi}_0 d\theta \sin^2\theta\,
 P.E.\left[
f^V(q) (e^{2i\theta} +e^{-2i\theta} +1) + f^{{1\over 2}H}(q) (e^{i\theta}+e^{-i\theta}) ( z+z^{-1})
\right]\, \nonumber\\
&= 1+  q + \left( -{1\over z^2} +2 -z^2 \right) q^2 + 
\left( -{1\over z^2} +2 -z^2\right)q^3 + \left( -{2\over z^2} +4- 2z^2\right)  q^4 \nonumber \\
&+  \left( {1\over z^4}  - {3\over z^2} +5 -3z^2 +z^4\right)q^5+
+\left( {1\over z^4}  -{5\over z^2} +8 -5z^2 +z^4 \right)q^6
+\cdots\,,
\end{align}
which agrees with  $\mathcal{I}_{SU(2),N_f=1} (q,z) = (q)_\infty^2 \,\text{Tr} [\mathcal{O}(q)]$ as a function of two variables.

\subsubsection{$SU(2)$ with $N_f=2$}

\begin{figure}[h]
\begin{center}
\includegraphics[width=.3\textwidth]{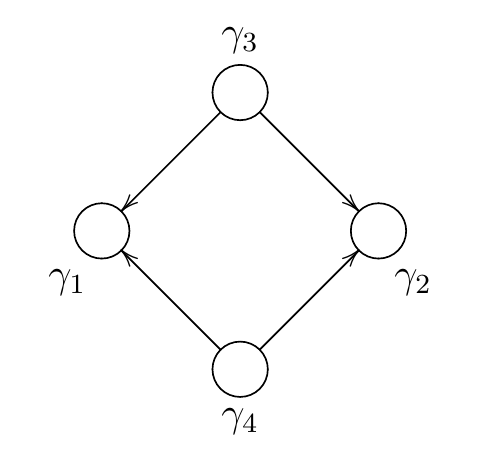}
\end{center}
\caption{ The BPS quiver for the $\mathcal{N}=2$ $SU(2)$ gauge theory with two hypermultiplets in the fundamental representation.}\label{fig:Nf2}
\end{figure}

The $\mathcal{N}=2$ $SU(2)$ gauge theory with two hypermultiplets in the fundamental representation has an $SO(4)$ flavor symmetry, whose fugacities will be denoted by $z_1,z_2$.\footnote{ Our convention for the character is, for example, $\chi_{\bf (2,2)}^{SO(4)}(z_1,z_2)= (z_1+z_1^{-1})(z_2+z_2^{-1})$. } 
 The BPS quiver is shown in Figure \ref{fig:Nf2}. There is a strong coupling chamber where \begin{align}
\text{arg}\, \mathcal{Z}(\gamma_1)=\text{arg}\, \mathcal{Z}(\gamma_2)>\text{arg}\, \mathcal{Z}(\gamma_3)=\text{arg}\, \mathcal{Z}(\gamma_4)\,.
\end{align}
 In this chamber  the only BPS particles are the nodes $\gamma_1$, $\gamma_2$, $\gamma_3$, $\gamma_4$. The lattice vectors for the  flavor symmetry are
 \begin{equation}
 \gamma_{f_1} = \gamma_1-\gamma_2~, \hspace{.5in}\gamma_{f_2} =\gamma_3-\gamma_4~.  \label{Nf=2flavor}
 \end{equation}
We will normalize the trace of the flavor generators to be
\begin{align}
\text{Tr}[X_{\gamma_{f_i}}] = z_i^2,~~~i=1,2\,.
\end{align}

The trace of the KS operator can be similarly evaluated to be
\begin{align}
\text{Tr}[\mathcal{O}(q)] & =  \sum_{\substack{\ell_1,\cdots,\ell_4,\\ k_1,\cdots, k_4=0}}^\infty
{  (-1)^{\sum_{i=1}^4 (k_i+\ell_i)}  q^{ {1\over2} \sum_{i=1}^4 (k_i+\ell_i) } \over \prod_{i=1}^4 (q)_{k_i} (q)_{\ell_i}}
\text{Tr}\left[ 
X_{-\gamma_4}^{k_4} X_{-\gamma_3}^{k_3}   X_{-\gamma_2}^{k_2} X_{-\gamma_1}^{k_1} 
X_{\gamma_4}^{\ell_4} X_{\gamma_3}^{\ell_3}   X_{\gamma_2}^{\ell_2} X_{\gamma_1}^{\ell_1} 
\right]\\
&=  \sum_{\substack{\ell_1,\cdots,\ell_4,\\ k_1,\cdots, k_4=0}}^\infty
{  (-1)^{\sum_{i=1}^4 (k_i+\ell_i)}  q^{ {1\over2} \sum_{i=1}^4 (k_i+\ell_i) +(k_1+k_2)(\ell_3+\ell_4) } \over \prod_{i=1}^4 (q)_{k_i} (q)_{\ell_i}}
\text{Tr}\left[ 
\prod_{i=1}^4 X_{\gamma_i}^{-k_i+\ell_i}
\right]\,. \nonumber
\end{align}
Now using \eqref{Nf=2flavor}, we can perform the following replacements,
\begin{equation}
X_{\gamma_1}  = X_{\gamma_2}X_{\gamma_{f_1}}~, \hspace{.5in} X_{\gamma_3}  = X_{\gamma_4}X_{\gamma_{f_2}}~.
\end{equation}
and obtain the final expression for the trace of the KS operator,
\begin{align}
\begin{split}
\text{Tr}[\mathcal{O}(q)] & =
\sum_{\substack{\ell_1,\cdots,\ell_4,\\ k_1,\cdots, k_4=0}}^\infty
{  q^{ \sum_{i=1}^4 \ell_i +(\ell_1+\ell_2)(\ell_3+\ell_4) } \over \prod_{i=1}^4 (q)_{k_i} (q)_{\ell_i}}
z_1^{2(\ell_1-k_1 )}z_2^{2(\ell_3-k_3)} \, \delta_{k_1+k_2,\ell_1+\ell_2} \delta_{k_3+k_4,\ell_3+\ell_4}\,.
\end{split}
\end{align}
We therefore obtain a $q$-expansion for
\begin{align}
\begin{split}
&(q)_\infty^2 \text{Tr}[\mathcal{O}(q)]  =
1+\left(  \chi_{(\bf 3,1)}^{SO(4)} +\chi_{\bf (1,3)}^{SO(4)}\right) \,q 
+\left(\chi_{ \bf (5,1)}^{SO(4) }  +\chi_{\bf  (3,1)}^{SO(4) } +\chi_{\bf (1,1)}^{SO(4) }+\chi_{\bf (1,3)}^{SO(4) }+\chi_{\bf (1,5) }^{SO(4) } \right) \, q^2\\
&+\left(\chi_{\bf (7,1)}^{SO(4) } +\chi_{\bf (5,1)}^{SO(4) } +2\chi_{\bf (3,1)}^{SO(4) }
+\chi_{\bf (1,1)}^{SO(4) } +2\chi_{\bf (1,3)}^{SO(4) }+\chi_{\bf (1,5)}^{SO(4) }+ \chi_{\bf (1,7)}^{SO(4) }  +\chi_{\bf (3,3)} ^{SO(4)}\right)q^3+\cdots\,,
\end{split}
\end{align}
where $\chi_{R}^{SO(4)}(z_1,z_2)$ is the character of $SO(4)$ in the representation $R$.

Meanwhile, the index for the $SU(2)$ gauge theory with two flavors is
\begin{align}
&\mathcal{I}_{SU(2),N_f=2} (q,z_1,z_2)  = {1\over \pi} \int^{2\pi}_0 d\theta \sin^2\theta\,
\nonumber \\&
\times
P.E.\left[
f^V(q) (e^{2i\theta} +e^{-2i\theta} +1) + f^{{1\over 2}H}(q) (e^{i\theta}+e^{-i\theta})\chi_{\bf (2,2)}^{SO(4)}(z_1,z_2)\right]\, \\
&=1+\left(  \chi_{(\bf 3,1)}^{SO(4)} +\chi_{\bf (1,3)}^{SO(4)}\right) \,q 
+\left(\chi_{ \bf (5,1)}^{SO(4) }  +\chi_{\bf  (3,1)}^{SO(4) } +\chi_{\bf (1,1)}^{SO(4) }+\chi_{\bf (1,3)}^{SO(4) }+\chi_{\bf (1,5) }^{SO(4) } \right) \, q^2 \nonumber\\
&+\left(\chi_{\bf (7,1)}^{SO(4) } +\chi_{\bf (5,1)}^{SO(4) } +2\chi_{\bf (3,1)}^{SO(4) }
+\chi_{\bf (1,1)}^{SO(4) } +2\chi_{\bf (1,3)}^{SO(4) }+\chi_{\bf (1,5)}^{SO(4) }+ \chi_{\bf (1,7)}^{SO(4) }  +\chi_{\bf (3,3)} ^{SO(4)}\right)q^3+\cdots\,~. \nonumber
\end{align}
This agrees with $\mathcal{I}_{SU(2),N_f=2} (q,z_1,z_2) = (q)_\infty^2 \text{Tr} [\mathcal{O}(q)]$  as a function of three variables.

\subsubsection{$SU(2)$ with $N_f=3$}

\begin{figure}[h]
\begin{center}
\includegraphics[width=.3\textwidth]{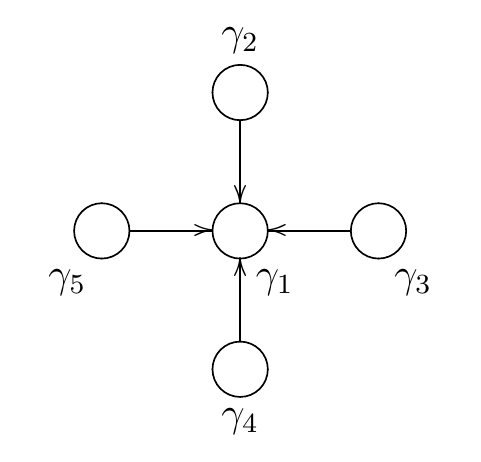}
\end{center}
\caption{ The BPS quiver for the $\mathcal{N}=2$ $SU(2)$ gauge theory with three hypermultiplets in the fundamental representation.}\label{fig:Nf3}
\end{figure}

The $\mathcal{N}=2$ $SU(2)$ gauge theory with three hypermultiplets in the fundamental representation has an $SO(6)$ flavor symmetry, whose fugacities will be denoted by $z_1,z_2,z_3$.\footnote{Our convention for the $SO(6)$ character is, for example, \begin{equation}\chi_{\bf 6}^{SO(6)} (z_1,z_2,z_3) =  z_2 + z_1z_2^{-1} z_3 +z_1^{-1}z_3  + z_1z_3^{-1}  +z_1^{-1} z_2z_3^{-1} +z_2^{-1}~.\end{equation}} The BPS quiver is shown in Figure \ref{fig:Nf3}. There is a strong coupling chamber where
\begin{equation}
 \text{arg}\, \mathcal{Z}(\gamma_1)>\text{arg}\, \mathcal{Z}(\gamma_2)=\text{arg}\, \mathcal{Z}(\gamma_3)=\text{arg}\, \mathcal{Z}(\gamma_4)=\text{arg}\, \mathcal{Z}(\gamma_5)~. 
 \end{equation}
In this chamber  the only BPS particles are the nodes $\gamma_1$, $\gamma_2$, $\gamma_3$, $\gamma_4$, $\gamma_5$.  We will choose the basis for the flavor symmetry to be 
\begin{equation}
\gamma_{f_1} ={1\over4} \left(- \gamma_2 -\gamma_3 -\gamma_4+3\gamma_5\right)~, \hspace{.05in}\gamma_{f_2} ={1\over2} \left( \gamma_2 -\gamma_3 -\gamma_4+\gamma_5\right)~, \hspace{.05in}\gamma_{f_3} ={1\over4} \left( \gamma_2 +\gamma_3 -3\gamma_4+\gamma_5\right)~,\label{Nf=3flavor}
\end{equation}
and normalize the trace of the flavor generators to be
\begin{align}
\text{Tr}[X_{\gamma_{f_i} }] =z_i\,,~~~i=1,2,3\,.
\end{align}
This  basis will turn out to be the most convenient choice when comparing with the index.

The trace of the KS operator can be similarly computed to be
\begin{align}
\begin{split}
\text{Tr}[\mathcal{O}(q)] 
&=  \sum_{\substack{\ell_1,\cdots,\ell_5,\\ k_1,\cdots, k_5=0}}^\infty
{  (-1)^{\sum_{i=1}^5 (k_i+\ell_i)}  q^{ {1\over2} \sum_{i=1}^5 (k_i+\ell_i) + k_1\sum_{i=2}^5 \ell_i } \over \prod_{i=1}^5 (q)_{k_i} (q)_{\ell_i}}
\text{Tr}\left[ 
\prod_{i=1}^5 X_{\gamma_i}^{-k_i+\ell_i}
\right]\,.
\end{split}
\end{align}
Now using \eqref{Nf=3flavor}, we can perform the following replacements,
\begin{equation}
X_{\gamma_2}  = X_{\gamma_5}X_{\gamma_{f_1}}^{-2} X_{\gamma_{f_2}}~, \hspace{.4in}X_{\gamma_3}  = X_{\gamma_5}X_{\gamma_{f_1}}^{-1}X_{\gamma_{f_2}}^{-1}X_{\gamma_{f_3}}~,\hspace{.4in}X_{\gamma_4}  = X_{\gamma_5}X_{\gamma_{f_1}}^{-1}X_{\gamma_{f_3}}^{-1}~,
\end{equation}
and obtain the final expression for the trace of the KS operator,
\begin{align}
\begin{split}
&\text{Tr}[\mathcal{O}(q)] =  \sum_{\substack{\ell_1,\cdots,\ell_5,\\ k_1,\cdots, k_5=0}}^\infty
{    q^{ \sum_{i=1}^5 \ell_i + \ell_1\sum_{i=2}^5 \ell_i } \over \prod_{i=1}^5 (q)_{k_i} (q)_{\ell_i}} 
z_1^{-(\ell_2-k_2) +(\ell_5-k_5)}\\
&~~~~~~~~~~~~~~~~~~~~~~~~~~
\times
z_2^{( \ell_2-k_2) -(\ell_3-k_3 )}
z_3^{( \ell_3-k_3)-(\ell_4-k_4)}
\,\delta_{k_1,\delta_1}\delta_{\sum_{i=2}^5 k_i , \sum_{i=2}^5 \ell_i} \, .
\end{split}
\end{align}
We therefore obtain a $q$-expansion for
\begin{align}
\begin{split}
(q)_\infty^2\text{Tr} [\mathcal{O}(q) ] 
&= 1+  \chi_{\bf 15  }^{SO(6)}(z_1,z_2,z_3)q+
\left( \chi_{\bf 1}^{SO(6)}  + \chi_{\bf 15}^{SO(6)} + \chi_{\bf 84} ^{SO(6)}  \right)(z_1,z_2,z_3)\,q^2+\cdots\,.
\end{split}
\end{align}

Meanwhile, the index for the $SU(2)$ gauge theory with two flavors is
\begin{align}
&\mathcal{I}_{SU(2),N_f=3} (q,z_1,z_2,z_3) \nonumber \\=
& {1\over \pi} \int^{2\pi}_0 d\theta \sin^2\theta\,
P.E.\left[
f^V(q) (e^{2i\theta} +e^{-2i\theta}+1 ) + f^{{1\over 2}H}(q) (e^{i\theta}+e^{-i\theta}) \chi_{\bf 6}^{SO(6)}(z_1,z_2,z_3)\right]\, \nonumber\\
&= 1+  \chi_{\bf 15  }^{SO(6)}(z_1,z_2,z_3)q+
\left( \chi_{\bf 1}^{SO(6)}  + \chi_{\bf 15}^{SO(6)} + \chi_{\bf 84} ^{SO(6)}  \right)(z_1,z_2,z_3)\,q^2+\cdots\,, 
\end{align}
which agrees with  $\mathcal{I}_{SU(2),N_f=3} (q,z_1,z_2,z_{3}) = (q)_\infty^2 \, \text{Tr} [\mathcal{O}(q)]$ as a function of four variables.

\section{Applications to Argyres-Douglas Theories}
\label{AD}

In this section, we assume the validity of our conjecture \eqref{conjflav} and apply it to compute the index of the strongly-coupled Argyres-Douglas conformal field theories \cite{Argyres:1995jj,Argyres:1995xn}.  These theories are labelled by and ADE singularity plane curve singularity which describes the (singular) Seiberg-Witten curve at the conformal fixed point.

The BPS spectra of the Argyres-Douglas theories is known \cite{Shapere:1999xr,Cecotti:2010fi,Gaiotto:2010be,Cecotti:2011rv,Alim:2011ae,Alim:2011kw}. As discovered in \cite{Cecotti:2010fi}, for all models, there is a chamber where the only BPS particles can be represented as a node of the associated ADE quiver diagram and every node is either a sink (all arrows coming in) or a source (all arrows going out). We will refer this to the sink/source chamber. The only BPS particles in the sink/source chamber are those corresponding to the nodes with the following phase order
\begin{align}
\text{arg}\, \mathcal{Z}(\gamma_I ) > \text{arg}\,\mathcal{Z}(\gamma_J)\,,~~~~~~\text{for all}~ I\in\text{sink},~J\in \text{source}\,.
\end{align}
We will use the indices $I, \,I'$ for the sink nodes and $J,\,J'$ for the source nodes. 

 The KS operator can then be written uniformly as \cite{Cecotti:2010fi}
\begin{align}\label{ssKS}
\mathcal{O}(q) = 
\prod_{J'\in \text{source}} E_q( X_{-\gamma_{J'}})   \prod_{I'\in \text{sink}} E_q( X_{-\gamma_{I'}})  
\prod_{J\in \text{source}} E_q( X_{\gamma_{J}})   \prod_{I\in\text{ sink}} E_q( X_{\gamma_{I}})  \,.
\end{align}
We evaluate the trace of this operator and use it to predict the index.

A strong check on our results comes from recent work connecting chiral algebras to four-dimensional $\mathcal{N}=2$ theories \cite{Beem:2013sza,Beem:2014rza,Lemos:2014lua}.  Specifically, it is known that the operators that may contribute to the Schur index form the structure of a two-dimensional chiral algebra.  Moreover the central charges of this chiral algebra are inherited from four dimensional central charges as \cite{Beem:2013sza}
\begin{equation}
c_{2d}=-12c_{4d}~, \hspace{.5in}k_{2d}=-\frac{1}{2}k_{4d}~, \label{2d4d}
\end{equation}
where $c_{2d}$ is the two-dimensional central charge, and $k_{2d}$ is the level of a flavor symmetry (if any is present).  Based on these results, one expects the Schur index to be modular with the given $c_{2d}.$  Since the central charges $c_{4d}$ of the Argyres-Douglas theories are known, this provides a prediction for our calculations.  Moreover, as anticipated from the work of \cite{Cecotti:2010fi} the trace of $\mathcal{O}(q)$ should also be related to a non-unitary minimal model.

Our results match perfectly with these expectations.  In the series $A_{2n},$  $D_{2n+1},$ $D_4$, and $E_{6}$ and $E_{8}$ we find characters of non-unitary minimal models or Kac-Moody algebras with known $2d$ central charges exactly agreeing with \eqref{2d4d}. Interestingly in the $A_{2n}$ case, the $q$ series we produce appear similar to the fermionic representation of Virasoro minimal model characters appearing in \cite{Feigin:1991wv, Berkovich:1994es}.

As another consistency condition, we observe that in the case of the $A_{2n+1}$ and $D_{2n}$  Argyres-Douglas theories, our formulas agree with the recent results of \cite{Buican:2015ina} derived by different reasoning.

Bolstered by these calculations, we conclude in \S\ref{sec:generalAD} with a proposal for the generalized Argyres-Douglas theories of type $(A_{k-1},A_{N-1}).$  Specifically, we conjecture that for $k$ and $N$ coprime, the chiral algebra of this theory is the vacuum sector of the non-unitary $W_{k}$ $(k,k+N)$ minimal model, and that the Schur index is the associated vacuum character. 

\subsection{$A_{n}$ Theories}

\subsubsection{$A_{2n}$}\label{sec:A2n}

\begin{figure}[h]
\begin{center}
\includegraphics[width=.8\textwidth]{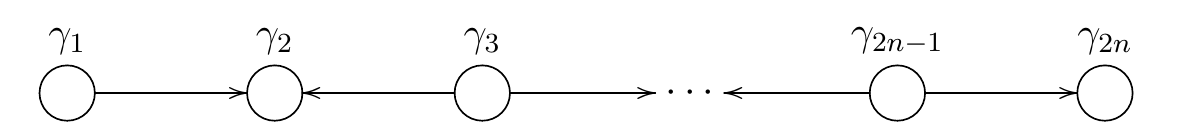}
\end{center}
\caption{ The BPS quiver for the $A_{2n}$ Argyres-Douglas theory in the sink/source chamber.}\label{fig:A2n}
\end{figure}

For the $A_{2n}$ Argyres-Douglas theory, the 4$d$ central charge is $c_{4d}={n(6n+5)\over 6(2n+3)}$ \cite{Aharony:2007dj,Shapere:2008zf,Xie:2012hs,Xie:2013jc}.  There is no flavor symmetry.  From the 4$d$/2$d$ relation $c_{2d}=-12c_{4d}$, we determine the 2$d$ central charge of the chiral algebra to be 
\begin{align}\label{c2dA2n}
c_{2d}  =- { 2n (6n+5)\over 2n+3}\,.
\end{align}
On the other hand, for two coprime integers $p$ and $p'$, the central charge of the $(p,p')$ Virasoro minimal model is given by
\begin{align}\label{W2c}
c^{(2)}(p,p')  = 1 - 6{(p-p')^2\over pp'}\,,
\end{align}
where the superscript (2) indicates that this is a minimal model of the Virasoro algebra. 
We recognize $c_{2d}$ in \eqref{c2dA2n} as the central charge  of the $(p=2,p'= 2n+3)$ Virasoro minimal model. 

Indeed, we will see that the Schur index of the $A_{2n}$ Argyres-Douglas theory is the vacuum character of the $(2,2n+3)$ Virasoro minimal model as conjectured by \cite{Lecture1, Lecture2}.  This is an extension of the calculations of \cite{Cecotti:2010fi} where it was found that, in the case of the $A_{2}$ model, the trace of $\mathcal{O}(q)^{-1}$ is closely related to the character of the non-trivial primary in the $(2,5)$ Virasoro minimal model, and confirms their more general expectation of a link between traces of powers of $\mathcal{O}(q)$ and characters of non-unitary Virasoro minimal models.

We will work in the sink/source chamber where all the even (odd) nodes are sinks (sources). The KS operator $\mathcal{O}(q)$ is then given by \eqref{ssKS}. By expanding out $E_q(x)$ as in \eqref{qexp}, the trace of the KS operator can be written as
\begin{align}
\begin{split}
&\text{Tr}[\mathcal{O}(q)]\\
&
= \sum_{\substack{\ell_1,\cdots,\ell_{2n},\\ k_1,\cdots ,k_{2n}=0} }^\infty
{(-1)^{\sum_{i=1}^{2n} (k_i+\ell_i) } q^{{1\over2}  \sum_{i=1}^{2n} (k_i+\ell_i ) } \over\prod_{i=1}^{2n} (q)_{k_i} (q)_{\ell_i}}
\text{Tr} \left[
\prod_{J'\in \text{odd}} X_{-\gamma_{J'}}^{k_{J'}}   \prod_{I'\in \text{even}}  X_{-\gamma_{I'}}^{k_{I'}} 
\prod_{J\in \text{odd}} X_{\gamma_{J}}^{\ell_J}  \prod_{I\in\text{even}} X_{\gamma_{I}}^{\ell_I}
\right]~.
\end{split}
\end{align}
By passing  $\prod_{I'\in \text{even}} X_{-\gamma_{I'}}^{k_{I'}}$ through $\prod_{J\in \text{odd}}  X_{\gamma_{J}}^{\ell_J} $, we pick up a power of $q$ from the quantum torus algebra \eqref{qtorus},
\begin{align}
\prod_{I'\in \text{even}} X_{-\gamma_{I'}}^{k_{I'}}\prod_{J\in \text{odd}}  X_{\gamma_{J}}^{\ell_J} 
=q^{ \,   \ell_1k_2 + \ell_3(k_2+k_4) +\cdots \ell_{2n-1}(k_{2n-2} +k_{2n})}\prod_{J\in \text{odd}}  X_{\gamma_{J}}^{\ell_J} \prod_{I'\in \text{even}} X_{-\gamma_{I'}}^{I'}\,.
\end{align}
It follows that
\begin{align}
\begin{split}
\text{Tr}[\mathcal{O}(q)]&
= \sum_{\substack{\ell_1,\cdots,\ell_{2n},\\ k_1,\cdots ,k_{2n}=0} }^\infty
{(-1)^{\sum_{i=1}^{2n} (k_i+\ell_i) } q^{{1\over2}  \sum_{i=1}^{2n} (k_i+\ell_i ) }q^{ \,  \ell_1k_2 + \ell_3(k_2+k_4) +\cdots \ell_{2n-1}(k_{2n-2} +k_{2n})} \over\prod_{i=1}^{2n} (q)_{k_i} (q)_{\ell_i}}\\
&\times
\text{Tr} \left[
\prod_{J\in \text{odd}} X_{\gamma_{J}}^{\ell_J-k_{J}}   \prod_{I\in\text{even}} X_{\gamma_{I}}^{\ell_I-k_I}
\right]~.
\end{split}
\end{align}
Finally, since there is no flavor symmetry, we have Tr$[X_{\gamma_i}]=0$ for all $i=1,\cdots,2n$. The trace then enforces a Kronecker delta $\delta_{k_i,\ell_i}$ associated to each node. Assuming our conjecture $\mathcal{I}_{A_{2n}}(q)=  (q)_\infty^{2n}\text{Tr} [\mathcal{O}(q)]$, we arrive at our final formula for the Schur index of the $A_{2n}$ Argyres-Douglas theory
\begin{align}\label{A2nSchur}
\begin{split}
\mathcal{I}_{A_{2n}}(q)
= (q)_\infty^{2n}\sum_{\ell_1,\cdots,\ell_{2n}=0 }^\infty
{ q^{  \sum_{i=1}^{2n} \ell_i  \,+ \sum_{i=1}^{2n-1} \ell_i \ell_{i+1}} \over\prod_{i=1}^{2n} [(q)_{\ell_i}]^2 }\,.
\end{split}
\end{align}
It will be more illuminating to write the answer as
\begin{align}
\begin{split}
\mathcal{I}_{A_{2n}}(q)
= (q)_\infty^{2n}\sum_{\ell_1,\cdots,\ell_{2n}=0 }^\infty
{ q^{  \sum_{i=1}^{2n} \ell_i  \,+{1\over2} \sum_{i,j=1}^{2n} b^{A_{2n}}_{ij}\ell_i \ell_{j}} \over\prod_{i=1}^{2n} [(q)_{\ell_i}]^2 }\,,
\end{split}
\end{align}
where 
\begin{align}
b^{A_{2n}}_{ij} =  -C^{A_{2n}}_{ij}+2 \delta_{ij}\,,
\end{align}
with $C^{A_{2n}}_{ij}$ the Cartan matrix of $A_{2n}$. We will see similar expressions for other Argyres-Douglas theories.

 On the other hand, the vacuum character of the $(p,p')$ Virasoro minimal model is given by (see, for example, \cite{philippe1997conformal})\footnote{We normalize the character such that the leading term is 1.}
\begin{align}
\chi^{(p,p')}_0(q)  = q^{ {1\over 24}(c^{(2)}(p,p')-1)  } {1\over (q)_\infty} \sum_{\ell\in \mathbb{Z}}\left(
q^{(2p p' \ell +p -p')^2\over 4p p'}-q^{(2p p' \ell +p +p')^2\over 4p p'}
\right)\,,
\end{align}
with $c^{(2)}(p,p') = 1- 6{(p-p')^2\over p p'}$.

We conjecture that the Schur index $\mathcal{I}_{A_{2n}}(q)$ of the $A_{2n}$ Argyres-Douglas theory equals to the vacuum character $\chi^{(2,2n+3)}_0(q)$ of the $(2,2n+3)$ Virasoro minimal model. Let us examine this conjecture in the $A_2$ theory. The Schur index has the following $q$-expansion,
\begin{align}
\mathcal{I}_{A_{2}}(q)&= (q)_\infty^{2}\sum_{\ell_1,\ell_{2}=0 }^\infty
{ q^{ \ell_1+\ell_2  +\ell_1 \ell_{2}} \over [(q)_{\ell_1}(q)_{\ell_2}]^2 }\\
&
=1+0q + q^2+q^3+q^4 +q^5 + 2q^6+ 2q^7 + 3q^8 +3q^9 + 4q^{10} +4q^{11} +6q^{12}+\cdots\,. \nonumber
\end{align}
On the other hand, the (2,5) Virasoro minimal model vacuum character can be expanded as
\begin{align}
\chi_0^{(2,5)} (q)   &=  q^{-{27\over 120} }{1\over (q)_\infty} \sum_{\ell\in \mathbb{Z}} 
\left(
q^{ (20\ell-3)^2\over 40 }-q^{ (20\ell+7)^2\over 40 }
\right)\\
&=1+0q + q^2+q^3+q^4 +q^5 + 2q^6+ 2q^7 + 3q^8 +3q^9 + 4q^{10} +4q^{11} +6q^{12}+\cdots\,. \nonumber
\end{align}
Incidentally, the vacuum character of the (2,5) Virasoro minimal model is known to be equal to the Rogers-Ramanujan function $H(q):= \sum_{\ell=0}^\infty {q^{\ell^2+\ell}\over (q)_\ell }$.

The Schur index for the $A_4$ Argyres-Douglas theory is
\begin{align}
\mathcal{I}_{A_{4}}(q)&= (q)_\infty^{4}\sum_{\ell_1,\ell_{2},\ell_3,\ell_4=0 }^\infty
{ q^{ \ell_1+\ell_2  +\ell_3+\ell_4+\ell_1 \ell_{2}+\ell_2\ell_3 +\ell_3\ell_4} \over [(q)_{\ell_1}(q)_{\ell_2}(q)_{\ell_3}(q)_{\ell_4}]^2 }\\
&= 1+0q+ q^2+q^3 + 2q^4+2q^5 + 3q^6  + 3q^7+5q^8+ 6q^9 +8q^{10}+ 9q^{11}+13q^{12}+\cdots\,. \nonumber
\end{align}
On the other hand, the vacuum character for the (2,7) Virasoro minimal model is
\begin{align}
\chi^{(2,7)}_0 (q) &= q^{ -{75\over 168} } {1\over (q)_\infty} \sum_{\ell\in\mathbb{Z}} \left(
q^{ (28\ell -5 )^2\over 56 }-q^{ (28\ell +9 )^2\over 56 }
\right)\\
&= 1+0q+ q^2+q^3 + 2q^4+2q^5 + 3q^6  + 3q^7+5q^8+ 6q^9 +8q^{10}+ 9q^{11}+13q^{12}+\cdots\,. \nonumber
\end{align}

As a further consistency check, we notice that the first few terms in the  vacuum character of a Virasoro minimal model  have the following universal form,
\begin{align}
\chi^{(2,2n+3)}_0 = 1+ 0q+q^2+\cdots\,,
\end{align}
which has the desired property for the Schur index of the $A_{2n}$ Argyres-Douglas theory. Indeed, as discussed in \eqref{shurexp}, the vanishing of the linear coefficient implies the absence of any flavor symmetry, and the 1 for the $q^2$ term is naturally interpreted as the unique energy-momentum tensor multiplet.

\subsubsection{$A_{2n+1}$}

\begin{figure}[h]
\begin{center}
\includegraphics[width=.8\textwidth]{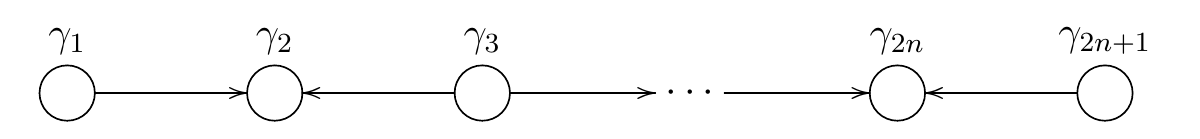}
\end{center}
\caption{ The BPS quiver for the $A_{2n+1}$ Argyres-Douglas theory in the sink/source chamber.}\label{fig:A2n1}
\end{figure}

For the $A_{2n+1}$ Argyres-Douglas theory, the 4$d$ central charge is $c_{4d}= {3n^2+5n+1\over 6(n+2)}$  \cite{Aharony:2007dj,Shapere:2008zf,Xie:2012hs,Xie:2013jc}. 
 There is an $U(1)$ flavor symmetry if $n>1$ and an $SU(2)$ flavor symmetry if $n=1$.  
 From the 4$d$/2$d$ relation $c_{2d}=-12c_{4d}$, we determine the 2$d$ central charge of the chiral algebra to be 
\begin{align}\label{c2dA2n1}
c_{2d}  =- { 2(3n^2+5n+1)\over n+2}\,.
\end{align}

We will choose the lattice vector $\gamma_f$ for the flavor symmetry to be
\begin{align}\label{A2n1flavor}
\gamma_f  =  \sum_{i=0}^n (-1)^i \gamma_{2i+1}\,.
\end{align}
We will normalize the trace of the flavor generator to be
\begin{align}
\text{Tr}[X_{\gamma_f}] =
\begin{cases}
&z^2\,,~~~~~~~~~~~~~~\text{if}~~n=1\,,\\
& (-1)^{n+1} z\,,~~~~~\text{if}~~n>1\,.
\end{cases}
\end{align}

Again in the sink/source chamber, the KS operator is given by \eqref{ssKS}.  Following an identical calculation as the $A_{2n}$ case, the trace of the KS operator can be written as
\begin{align}
\text{Tr}[\mathcal{O}(q)]&=  \sum_{\substack{\ell_1,\cdots,\ell_{2n+1},\\ k_1,\cdots ,k_{2n+1}=0} }^\infty
{(-1)^{\sum_{i=1}^{2n+1} (k_i+\ell_i) } q^{{1\over2}  \sum_{i=1}^{2n+1} (k_i+\ell_i ) }q^{ \,  k_2(\ell_1+\ell_3) + k_4(\ell_3+\ell_5) +\cdots k_{2n}(\ell_{2n-1} +\ell_{2n+1})} \over\prod_{i=1}^{2n+1} (q)_{k_i} (q)_{\ell_i}}  \nonumber \\
&\times
\text{Tr} \left[
\prod_{J\in \text{odd}} X_{\gamma_{J}}^{\ell_J-k_{J}}   \prod_{I\in\text{even}} X_{\gamma_{I}}^{\ell_I-k_I}
\right]\,.
\end{align}
Using \eqref{A2n1flavor}, we can perform the following replacement
\begin{align}\label{A2n1replace}
X_{\gamma_{1} } =  X_{\gamma_f} \prod_{i=1}^{n} (X_{\gamma_{2i+1}})^{(-1)^{i+1}}\,.
\end{align}
The traces on the even nodes then enforce the Kronecker delta $\delta_{k_{2i}, \ell_{2i}}$ while those on the odd nodes enforce $\delta_{(-1)^{j+1}k_1 + k_{2j+1} ,(-1)^{j+1}\ell_1 + \ell_{2j+1}  }$ (except for $\gamma_1$ which has already been solved for in terms of other charges in \eqref{A2n1replace}). 
Assuming our conjecture $\mathcal{I}_{A_{2n+1}}(q,z) =(q)_\infty^{2n}\text{Tr}[\mathcal{O}(q)]$, we obtain the Schur index for the $A_{2n+1}$ Argyres-Douglas theory
\begin{align}
\begin{split}
\mathcal{I}_{A_{2n+1}}(q,z) &=(q)_\infty^{2n} \sum_{\substack{\ell_1,\cdots,\ell_{2n+1},\\ k_1,\cdots ,k_{2n+1}=0} }^\infty
{(-1)^{\sum_{i=1}^{2n+1} (k_i+\ell_i) } q^{{1\over2}  \sum_{i=1}^{2n+1} (k_i+\ell_i ) + \sum_{j=1}^n  \ell_{2j}(\ell_{2j-1}+\ell_{2j+1}) } \over\prod_{i=1}^{2n+1} (q)_{k_i} (q)_{\ell_i}}\\
&\times\, \left[ (-1)^{n+1}z\right]^{\ell_1-k_1}\,  \prod_{i=1}^n \delta_{k_{2i},\ell_{2i}} \, \prod_{j=1}^n  \delta_{(-1)^{j+1}k_1 + k_{2j+1} ,(-1)^{j+1}\ell_1 + \ell_{2j+1}  }\,.
\end{split}
\end{align}

Because of the isomorphism $D_{3}\cong A_{3}$, the $A_3$ Argyres-Douglas theory is special in the $A_{2n+1}$ series and we defer its discussion to \S\ref{sec:D2n1} where we treat the whole series $D_{2n+1}$ uniformly. The $q$-expansion for the first few $A_{2n+1}$ theories with $n>1$ is
\begin{align}
&\mathcal{I}_{A_5} (q)  
=1+q +(z+z^{-1})  q^{3\over2}  + 3q^2 +(2z+2z^{-1})q^{5\over2}+
(z^2+5+z^{-2})q^3\notag  \\
&~~\,~~~~~~ + (4z+4z^{-1})q^{7\over2}  + (2z^2 +10 + 2z^{-2})q^4 + (z^3 +  8z+8z^{-1}+ z^{-3})q^{9\over2}
+\cdots~,\\
&\mathcal{I}_{A_7} (q)  
=1+q +(z+3+z^{-1})  q^{2}  + 
(2z+5+2z^{-1})q^3
+ ( z^2 + 4z + 10 +4z^{-1}+z^{-2} ) q^4
+\cdots\,, \nonumber\\
&\mathcal{I}_{A_9} (q)  
=1+q +3 q^{2}  + 
(z+z^{-1})q^{5\over2}+\cdots\,.~~~~~~~~~~~~~~~~~~~~~ \nonumber
\end{align}
Note that the linear term in $q$ encodes the correct $U(1)$ flavor symmetry. As a  consistency check, we note that the above $q$-expansions agree with the formula conjectured in \cite{Buican:2015ina}.

\subsection{$D_{n}$ Theories}

\subsubsection{$D_{2n+1}$}\label{sec:D2n1}

\begin{figure}[h]
\begin{center}
\includegraphics[width=.7\textwidth]{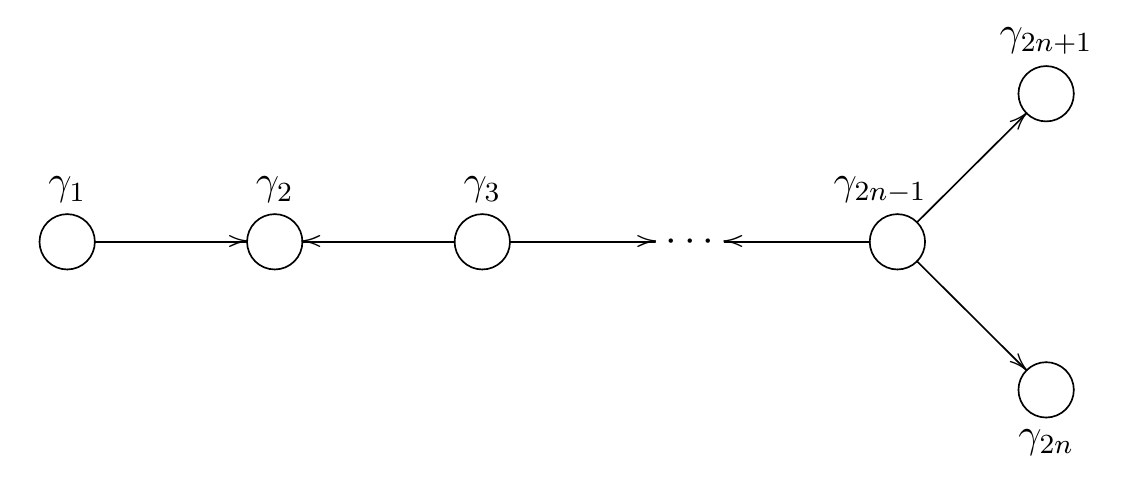}
\end{center}
\caption{ The BPS quiver for the $D_{2n+1}$ Argyres-Douglas theory in the sink/source chamber.}\label{fig:D2n1}
\end{figure}

The $D_{2n+1}$ theory has an $SU(2)$ flavor symmetry. The 4$d$ central charges $c_{4d} = {n\over2}$ and the flavor central charge $k_{4d} = { 8n\over 2n+1}$ \cite{Shapere:2008zf,Xie:2012hs,Xie:2013jc}. Using the 4$d$/2$d$ relations, $c_{2d} =-12 c_{4d}$ and $k_{2d} = -{1\over 2} k_{4d}$, we have the following 2$d$ central charge and level,
\begin{align}
c_{2d} = -6n,~~~~k_{2d} = - {4n\over 2n+1}\,.
\end{align}
Note that in this case the $SU(2)$ Sugawara central charge 
\begin{align}
D_{2n+1}:~~~~c_{\text{Sug}}  = {  3 k_{2d} \over k_{2d}  +2 }  =-6n= c_{2d}\,,
\end{align} 
saturates the 2$d$ central charge $c_{2d}$. This strongly suggests that the chiral algebra for the $D_{2n+1}$ Argyres-Douglas theory is the Kac-Moody algebra $\widehat{SU(2)}_{ - {4n\over 2n+1}}$.  Indeed, we will check that the Schur index of the $D_{2n+1}$ Argyres-Douglas theory equals to the vacuum character of $\widehat{SU(2)}_{ - {4n\over 2n+1}}$ as conjectured by \cite{Lecture1, Lecture2}.

We choose the flavor lattice vector to be
\begin{align}\label{D2n1flavor}
\gamma_f = \gamma_{2n+1} -\gamma_{2n}\,.
\end{align}
We normalize the trace of the flavor generator to be
\begin{align}
\text{Tr} [X_{\gamma_f} ] =z^2\,.
\end{align}

We again work in the sink/source chamber where $\gamma_2,\gamma_4,\cdots ,\gamma_{2n},\gamma_{2n+1}$ are sinks and $\gamma_1,\gamma_3,\cdots,\gamma_{2n-1}$ are sources. The KS operator is again given by \eqref{ssKS}.  By a similar calculation as in $A_{2n+1}$, the trace of the KS operator can be written as
\begin{align}
\begin{split}
\text{Tr}[\mathcal{O}(q) ]  & =  \sum_{\substack{ \ell_1,\cdots,\ell_{2n+1},\\ k_{2n},k_{2n+1}=0}}^\infty 
{  (-1)^{ k_{2n}+k_{2n+1} +\ell_{2n}+\ell_{2n+1}  }  q^{\sum_{i=1}^{2n-1}\ell_i  +   {1\over2}   (k_{2n}+k_{2n+1} +\ell_{2n}+\ell_{2n+1})   }\over (q)_{k_{2n}} (q)_{k_{2n+1}} (q)_{\ell_{2n}} (q)_{\ell_{2n+1}} \prod_{i=1}^{2n-1}[(q)_{\ell_i}]^2 }\\
&\times
q^{ \ell_{2n-1} (\ell_{2n-2} +k_{2n} +k_{2n+1}) +\sum_{i=1}^{n-1} \ell_{2i} (\ell_{2i-1} + \ell_{2i+1})  }
\, \text{Tr}\left[  
X_{\gamma_{2n}}^{-k_{2n} +\ell_{2n}}X_{\gamma_{2n+1}}^{-k_{2n+1} +\ell_{2n+1}}
\right]\,.
\end{split}
\end{align}
Note that we have already performed the trace on $X_{\gamma_i}$ which enforces $\ell_i=k_i$ for $i=1,\cdots,2n-1$. 
Using \eqref{D2n1flavor}, we can replace
\begin{align}
X_{\gamma_{2n+1}  }  = X_{\gamma_{2n} } X_{\gamma_f}  \,.
\end{align}
Assuming our conjecture $\mathcal{I}_{D_{2n+1}}(q) = (q)_\infty^{2n} \text{Tr} [\mathcal{O}(q) ]$, we arrive at our final expression for 
\begin{align}
\begin{split}
\mathcal{I}_{D_{2n+1} } (q,z)  &= 
(q)_\infty^{2n}
\sum_{\substack{ \ell_1,\cdots,\ell_{2n+1},\\ k_{2n},k_{2n+1}=0}}^\infty 
{  q^{\sum_{i=1}^{2n+1}\ell_i +    {1\over 2} \sum_{i,j=1}^{2n+1} b_{ij}^{D_{2n+1}}\ell_i \ell_j }\over (q)_{k_{2n}} (q)_{k_{2n+1}} (q)_{\ell_{2n}} (q)_{\ell_{2n+1}} \prod_{i=1}^{2n-1}[(q)_{\ell_i}]^2 }\\
&\times
 z^{2(\ell_{2n+1} -k_{2n+1}) } \, \delta_{k_{2n}+k_{2n+1} , \ell_{2n} + \ell_{2n+1}}\,,
\end{split}
\end{align} 
where $b_{ij}^{D_{2n+1}} = - C^{D_{2n+1} }_{ij}  +2\delta_{ij}$ and $C^{D_{2n+1} }_{ij}$ is the Cartan matrix of $D_{2n+1}$.

On the other hand, the vacuum character of $\widehat{SU}(2)_{-{4n\over 2n+1}}$ can be derived straightforwardly by applying the Kac-Wakimoto formula \cite{kac1988modular},
\begin{align}\label{su2character}
\begin{split}
&\chi_{\widehat{SU(2)}_{ -{4n\over 2n+1} }}(q,z)\\
&={1\over  \prod_{n=1}^\infty (1-q^n)(1-z^2 q^n) (1-z^{-2}q^n)}\sum_{m=0}^\infty (-1)^m \, 
{  z^{  2m+1}  -  z^{-(2m+1)} \over z-z^{-1} }  \,q^{  {m(m+1)\over 2} (2n+1)}\,.
\end{split}
\end{align}

The $q$-expansion for the Schur indices of the first two $D_{2n+1}$ theories are listed below.
\begin{align}
&\mathcal{I}_{D_3\cong A_3}  (q,z) = 1 + \chi_{\bf 3} ^{SU(2)} (z) q  + \left( \chi_{\bf 1}  ^{SU(2)} +\chi_{\bf3} ^{SU(2)} +\chi_{\bf5} ^{SU(2)} \right)(z) \, q^2  \notag\\
&~~~~~~~~~~~~~~~~+ 
\left( \chi_{\bf1}^{SU(2)}+2\chi_{\bf3} ^{SU(2)}+  \chi_{\bf 5} ^{SU(2)}+ \chi_{\bf 7} ^{SU(2)} \right)(z)\, q^3\notag\\
&~~~~~~~~~~~~~~~~+\left( 2\chi_{\bf1}^{SU(2)}+ 3\chi_{\bf 3}^{SU(2)}+  3 \chi_{\bf5}^{SU(2)}+ \chi_{\bf 7}^{SU(2)}+ \chi_{\bf 9} ^{SU(2)}\right)(z)\,q^4+\cdots\,,\\
&\mathcal{I}_{D_5} (q,z)  =  1+ \chi_{\bf 3} ^{SU(2)}(z) q + \left( \chi_{\bf 1} ^{SU(2)} +\chi_{\bf3}^{SU(2)} +\chi_{\bf5} ^{SU(2)}\right)(z) \, q^2\notag\\
&~~~~~~~~~~~~~~~~~~+
\left( \chi_{\bf1}^{SU(2)}+3\chi_{\bf3}^{SU(2)} +  \chi_{\bf 5}^{SU(2)} + \chi_{\bf 7}^{SU(2)}  \right)(z)\, q^3\notag\\
&~~~~~~~~~~~~~~~~~~+\left( 3\chi_{\bf1}^{SU(2)} + 4\chi_{\bf 3}^{SU(2)} +  4 \chi_{\bf5}^{SU(2)} + \chi_{\bf 7}^{SU(2)} + \chi_{\bf 9} ^{SU(2)} \right)(z)\,q^4+\cdots\,.
\end{align}
They match with the $q$-expansions for the vacuum characters of $\widehat{SU(2)}_{- {4n \over 2n+1}} $ \eqref{su2character} exactly. This strongly supports our claim,
\begin{align}
\mathcal{I}_{D_{2n+1}} (q,z) =  \chi_{\widehat{SU(2)}_{ -{4n\over 2n+1} }}(q,z)\,.
\end{align}

\subsubsection{$D_{2n+2}$}

\begin{figure}[h]
\begin{center}
\includegraphics[width=.7\textwidth]{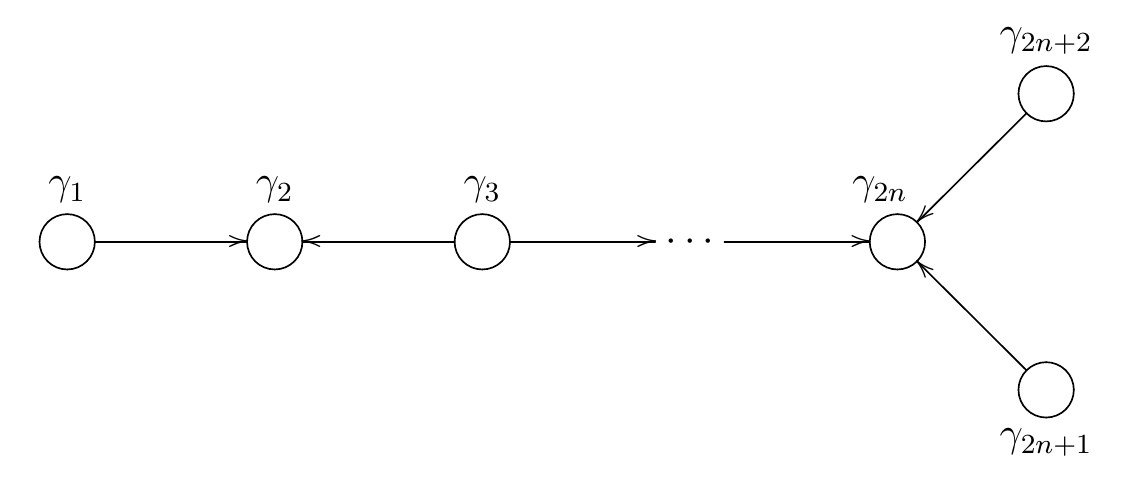}
\end{center}
\caption{ The BPS quiver for the $D_{2n+2}$ Argyres-Douglas theory in the sink/source chamber.}\label{fig:D2n2}
\end{figure}

The $D_{2n+2}$ Argyres-Douglas theory has an $SU(2)\times U(1)$ flavor symmetry for $n>1$ and $SU(3)$ flavor symmetry for $n=1$. The 4$d$ central charge and the flavor central charge are $c_{4d} = {n\over2} +{1\over 6}$ and $k_{4d} = {2  (2n+1)\over n+1}$ \cite{Aharony:2007dj,Shapere:2008zf,Xie:2012hs,Xie:2013jc}. Using the 4$d$/2$d$ relations $c_{2d} =-12 c_{4d}$ and $k_{2d} = -{1\over 2} k_{4d}$, we obtain
\begin{align}
c_{2d}=  -6n -2 \,,~~~~~~~k_{2d}  =  - {2n+1\over n+1}\,.
\end{align}
For the special case of $D_4$, the $SU(3)$ Sugawara central charge saturates the 2$d$ central charge,
\begin{align}
D_4:~~~c_{\text{Sug}} = { 8k_{2d} \over k_{2d} +3 } = -8 = c_{2d}\,.
\end{align}
Indeed, we will check that the Schur index of the $D_4$ Argyres-Douglas theory agrees with the vacuum character of $\widehat{SU(3)}_{-{3\over2}}$ as conjectured by \cite{Lecture1, Lecture2}.

We will choose flavor lattice vectors to be
\begin{equation}
\gamma_{f_1}  =  \gamma_{2n+2} -\gamma_{2n+1}~, \hspace{.5in}\gamma_{f_2} =  \sum_{i=0 }^n (-1)^i \gamma_{2i+1}~.  \label{D2n2flavor}
\end{equation}
We will use $y,x$ as variables for the characters of  $SU(2)\times U(1)$, with the convention $\chi_{\bf 2}^{SU(2)}(y) = y+y^{-1}$. The $D_4$ theory enjoys an enhanced $SU(3)$ flavor symmetry and we discuss the convention for that case in more detail later. To match with this convention for the characters,  we need to normalize the trace of the flavor generators in the following way, 
\begin{equation}
\text{Tr}[X_{\gamma_{f_1}} ]  =y^{2}~, \hspace{.5in}\text{Tr}[X_{\gamma_{f_2}} ]  = 
\begin{cases}
xy~~~~~~~~~~\text{if}~n~\text{is odd}\,,\\
-{x\over y}~~~~~~~~~\text{if}~n~\text{is even}\,.
\end{cases} \label{flavorrelation}
\end{equation}
Finally, using \eqref{D2n2flavor}, we can then replace the generators $X_{\gamma_1}$ and $X_{\gamma_{2n+2}}$ by
\begin{equation}
X_{\gamma_{2n+2}}  =  X_{\gamma_{f_1}}X_{\gamma_{2n+1}}~, \hspace{.5in}X_{\gamma_1} = X_{\gamma_{f_2} }  \prod_{i=1}^n (-1)^{i+1}  X_{\gamma_{2i+1}}~.
\end{equation}

After a similar calculation as for the $D_{2n+1}$ Argyres-Douglas theory, we obtain the Schur index of the $D_{2n+2}$ Argyres-Douglas theory,
\begin{align}
\mathcal{I}_{D_{2n+2} } (q,x,y)&  = 
(q)_\infty^{2n}
\sum_{\substack{ \ell_1,\cdots,\ell_{2n+2},\\ k_1,\cdots,k_{2n+2}=0}}^\infty 
{  (-1)^{\sum_{i=1}^{2n+2} (k_i +\ell_i)}  q^{{1\over2}\sum_{i=1}^{2n+2}( k_i+\ell_i )+    {1\over 2} \sum_{i,j=1}^{2n+2} b_{ij}^{D_{2n+2}}\ell_i\ell_j }\over  \prod_{i=1}^{2n+2}(q)_{k_i}(q)_{\ell_i} } \nonumber\\
&\times
 \left(\text{Tr} [X_{\gamma_{f_1}}]\right)^{\ell_{2n+2}  -k_{2n+2} }  
 \left(\text{Tr} [X_{\gamma_{f_2}}]\right)^{\ell_1-k_1} \,\left( \prod_{i=1}^{n}\delta_{ k_{2i} ,\ell_{2i}}\right)\\
  &\times
 \left( \prod_{i=1}^{n-1}  \delta_{ (-1)^{i+1} k_1+  k_{2i+1}  , (-1)^{i+1} \ell_1+  \ell_{2i+1} }\right)
 \delta_{ (-1)^{n+1}k_1 +k_{2n+1} +k_{2n+2} ,(-1)^{n+1}\ell_1 +\ell_{2n+1} +\ell_{2n+2}  }\,, \nonumber
\end{align}
where $b_{ij}^{D_{2n+2}} = -C^{D_{2n+2} }_{ij} +2\delta_{ij}$ with $C^{D_{2n+2} }_{ij}$ the Cartan matrix of $D_{2n+2}$. The traces of the flavor generators $\text{Tr} [X_{\gamma_{f_i}}]$ are given in \eqref{flavorrelation} as functions of the flavor fugacities $x,y$.

Let us start with the $D_4$ Argyres-Douglas theory. The $D_4$ theory enjoys an enhanced $SU(3)$ flavor symmetry so it would be more appropriate to change the flavor fugacity variables $x,y$ to the standard variables $z_1,z_2$ of the $SU(3)$ character.\footnote{ Our convention is such that $\chi_{\bf3}^{SU(3)} (z_1,z_2)=z_1 + z_1^{-1}z_2 +z_2^{-1}$.} The relation between $x,y$ and $z_1,z_2$ is
\begin{align}
\begin{split}
x=z_2^{-{3\over2}},~~~y=z_1z_2^{-{1\over2}}\,.
\end{split}
\end{align}
Equivalently, we have $\text{Tr} [X_{\gamma_{f_1}}] = y^2= z_1^2z_2^{-1}$ and $\text{Tr} [X_{\gamma_{f_2}}]=xy = z_1z_2^{-2}$.

The $q$-expansion of the  Schur index for the $D_4$ Argyres-Douglas theory is
\begin{align}
&\mathcal{I}_{D_4} (q,z_1,z_2)  =  1+  \chi^{SU(3)}_{\bf 8} \,q 
+\left( \chi^{SU(3)}_{\bf 1} + \chi^{SU(3)}_{\bf 8}+\chi^{SU(3)}_{\bf 27}\right) \,q^2  \nonumber \\
&
+\left( \chi^{SU(3)}_{\bf 1} + 2\chi^{SU(3)}_{\bf 8}+\chi^{SU(3)}_{\bf 10}+\chi^{SU(3)}_{\bf \overline{10}}+\chi^{SU(3)}_{\bf 27}+\chi^{SU(3)}_{\bf 64}\right)\,q^3\\
&+\left(2 \chi^{SU(3)}_{\bf 1} + 4\chi^{SU(3)}_{\bf 8}+\chi^{SU(3)}_{\bf 10}+\chi^{SU(3)}_{\bf \overline{10}}+3\chi^{SU(3)}_{\bf 27}+\chi^{SU(3)}_{\bf 35}+\chi^{SU(3)}_{\bf \overline{35}}+\chi^{SU(3)}_{\bf 64}+\chi^{SU(3)}_{\bf 125}\right) \,q^3+\cdots\,, \nonumber
\end{align}
where we have suppressed the explicit $z_1,z_2$ dependence on the $SU(3)$ characters.  This exactly matches the expansion of the vacuum character of $\widehat{SU(3)}_{ -{3\over2}}$, which can be computed using the Kac-Wakimoto formula \cite{kac1988modular} (see \cite{Buican:2015ina} for the explicit $q$-expansion).   This supports the claim
\begin{align}
\mathcal{I}_{D_4} (q,z_1,z_2)  = \chi_{\widehat{SU(3)}_{ -{3\over2}}} (q,z_1,z_2)\,.
\end{align}

For the other $D_{2n+2}$ theories, the flavor symmetries are $SU(2)\times U(1)$ and we record the $q$-expansion of their Schur indices below.
\begin{align}
&\mathcal{I}_{D_6} (q,x,y) = 
1+ \left(  \chi^{SU(2)}_{\bf1}
+ \chi^{SU(2)}_{\bf3}  \right) \,q
+(x+x^{-1} ) \chi^{SU(2)}_{\bf2} \,q^{3\over2}\notag\\
& ~~~~~~~~~~~
+\left(3\chi^{SU(2)}_{\bf1}+ 2\chi^{SU(2)}_{\bf3}+  \chi^{SU(2)}_{\bf5}\right)\,q^2
+(x+x^{-1}) \left(2\chi^{SU(2)}_{\bf2}+ \chi^{SU(2)}_{\bf4}\right)\,q^{5\over2}\notag\\
& ~~~~~~~~~~~+ \left[ \,5\chi^{SU(2)}_{\bf1}
+(x^2+6+x^{-2})\chi^{SU(2)}_{\bf3}
+2\chi^{SU(2)}_{\bf5}+
 \chi^{SU(2)}_{\bf7}\,\right]\,q^3+\cdots~,\\
 &\mathcal{I}_{D_8} (q,x,y) = 1+  \left(  \chi^{SU(2)}_{\bf1}
+ \chi^{SU(2)}_{\bf3} \right) \,q\notag\\
&~~~~~~~~
+\left[ 3\chi^{SU(2)}_{\bf1} +(x+x^{-1})\chi^{SU(2)}_{\bf2} +2\chi^{SU(2)}_{\bf3} +\chi^{SU(2)}_{\bf5} \right]\,q^2\notag\\
&~~~~~~~~
+\left[ 5\chi^{SU(2)}_{\bf1} +(x+x^{-1})\left( 2\chi^{SU(2)}_{\bf2} +\chi^{SU(2)}_{\bf4}\right)+6\chi^{SU(2)}_{\bf3}+2\chi^{SU(2)}_{\bf5} +\chi^{SU(2)}_{\bf7} \right]\,q^3+\cdots~, \nonumber
\end{align}
where we have suppressed the $y$-dependence on the $SU(2)$ characters. As a final consistency check, we note that the above $q$-expansions agree with the formula conjectured in \cite{Buican:2015ina}.

\subsection{$E_{n}$ Theories}

\subsubsection{$E_{6}$}\label{sec:E6}

\begin{figure}[h]
\begin{center}
\includegraphics[width=.6\textwidth]{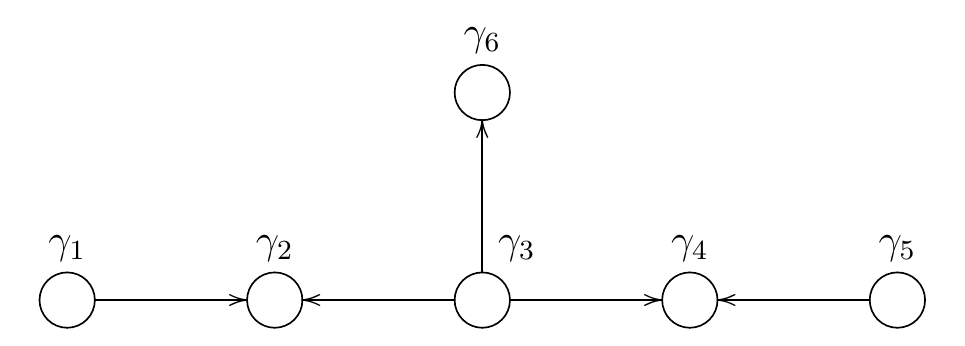}
\end{center}
\caption{ The BPS quiver for the $E_6$ Argyres-Douglas theory in the sink/source chamber.}\label{fig:E6}
\end{figure}

The $E_6$ Argyres-Douglas theory has no flavor symmetry. The 4$d$ central charge is $c_{4d}= {19\over14}$ \cite{Xie:2012hs,Xie:2013jc}. Using the 4$d$/2$d$ relation, we obtain the 2$d$ central charge,
\begin{align}
c_{2d} = - { 114\over 7}\,.
\end{align}
On the other hand, for coprime integers $p',p\ge3$,  the central charge of the $(p,p')$ minimal model of the $W_3$ algebra is \cite{Fateev:1987vh,Bouwknegt:1992wg}
\begin{align}\label{W3central}
c^{(3)}(p,p') = 2\left( 1- 12 {(p-p')^2\over p p' }\right)\,.
\end{align}
Note that $c_{2d}=-114/7$ is also the central charge of the (3,7) minimal model of $W_3$ algebra. Indeed we will show that the Schur index of the $E_6$ Argyres-Douglas theory equals to the vacuum character of the  (3,7) $W_3$ minimal model.

The trace of the KS operator can be computed in a similar way as the $A_{2n}$ case. Assuming the conjecture $\mathcal{I}_{E_6}(q) = (q)_\infty^6 \text{Tr} [\mathcal{O}(q)]$, we obtain the Schur index of the $E_6$ Argyres-Douglas theory,
\begin{align}\label{E6}
\mathcal{I}_{E_6}(q)&
= (q)_\infty^{6}\sum_{\ell_1,\cdots,\ell_{6}=0 }^\infty
{ q^{  \sum_{i=1}^{6} \ell_i  \,+{1\over2} \sum_{i,j=1}^{6} b^{E_6}_{ij}\ell_i \ell_{j}} \over\prod_{i=1}^{6} [(q)_{\ell_i}]^2 }\,,\\
&= 1+0q+q^2 + 2q^3 +3q^4 + 3q^5 + 6q^6 + 7q^7 + 11q^8  + 14q^9  +20q^{10} +25q^{11}+\cdots\,, \nonumber
\end{align}
where $b^{E_6}_{ij} =  -C^{E_6}_{ij}+2 \delta_{ij}$ and $C^{E_6}_{ij}$ is the Cartan matrix of $E_6$.

On the other hand, the vacuum character of the (3,7) $W_3$ minimal model is\cite{andrews1999a2} (see \S\ref{sec:generalAD} for more details)
\begin{align}
\chi^{W_3,\,(3,7)} (q) &= {[(q^7;q^7)_\infty]^2\over [(q;q)_\infty]^2 } [(q;q^7)_\infty ]^2(q^5;q^7)_\infty (q^2 ;q^7)_\infty[ (q^6; q^7)_\infty]^2\\
&= 1+0q+q^2 + 2q^3 +3q^4 + 3q^5 + 6q^6 + 7q^7 + 11q^8  + 14q^9  +20q^{10} +25q^{11}+\cdots\,, \nonumber
\end{align}
where the $q$-Pochhammer symbol $(a;q)_n$ is defined as
\begin{align}
(a;q)_n = \prod_{i=0}^{n-1}  (1-a q^k)\,. 
\end{align}
In particular, $(q;q)_n = (q)_n = \prod_{k=1}^n (1-q^k)$. The match of the $q$-expansion strongly supports our conjecture that the $E_6$ Schur index is the same as the vacuum character of the (3,7) $W_3$ minimal model,
\begin{align}
\mathcal{I}_{E_6} (q) = \chi^{W_3,\,(3,7)} (q) \,.
\end{align}

\subsubsection{$E_{7}$}

\begin{figure}[h]
\begin{center}
\includegraphics[width=.7\textwidth]{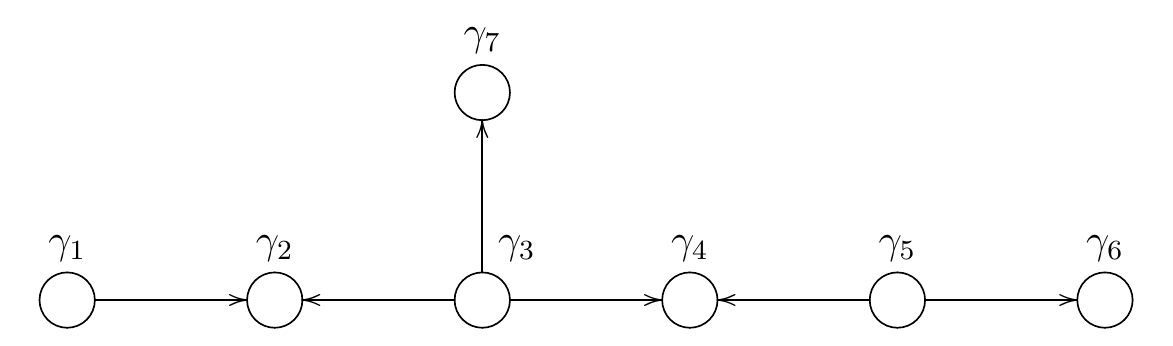}
\end{center}
\caption{ The BPS quiver for the $E_7$ Argyres-Douglas theory in the sink/source chamber.}\label{fig:E7}
\end{figure}

The $E_7$ Argyres-Douglas theory has an $U(1)$ flavor symmetry. The 4$d$ central charge is $c_{4d}= {31\over20}$ \cite{Xie:2012hs,Xie:2013jc}, which gives the 2$d$ central charge 
\begin{align}
c_{2d} =  - { 93\over 5}\,.
\end{align}
Choose the flavor lattice vector to be
\begin{align}
\gamma_f = \gamma_4 -\gamma_6 -\gamma_7\,.
\end{align}
We can then replace $X_{\gamma_7}$ by 
\begin{align}\label{E7replace}
X_{\gamma_7} =  X_{\gamma_f}^{-1}  X_{\gamma_4} X_{\gamma_6}^{-1}\,.
\end{align}
We will normalize the trace of the flavor generator to be
\begin{align}
\text{Tr} [X_{\gamma_f} ] = -z\,.
\end{align}

The trace of the KS operator can be similarly computed to be
\begin{align}
\text{Tr}[\mathcal{O}(q)] & = \sum_{\substack{\ell_1,\cdots, \ell_7, \\ k_1, \cdots , k_7=0}}^\infty
{(-1)^{\sum_{i=1}^7 (k_i +\ell_i ) }q^{ {1\over 2} \sum_{i=1}^7 (k_i +\ell_i )  }  
q^{k_2(\ell_1+\ell_3) +k_4 (\ell_3+\ell_5) +k_6\ell_5 +k_7\ell_3} 
\over \prod_{i=1}^7 (q)_{k_i} (q)_{\ell_i}}\\
&\times \text{Tr} \left[\left(\prod_{i=1}^3 X_{\gamma_{2i-1}}^{ -k_{2i-1}+\ell_{2i-1}}\right) X_{\gamma_7}^{-k_7+\ell_7}  
\left(\prod_{i=1}^3 X_{\gamma_{2i}}^{ -k_{2i}+\ell_{2i}}\right) 
\right]\,.
 \end{align}
Now using \eqref{E7replace} and assuming our conjecture $\mathcal{I}_{E_7} (q,z) = \text{Tr} [\mathcal{O}(q)]$, we obtain the Schur index of the $E_7$ Argyres-Douglas theory,
\begin{align}
\mathcal{I}_{E_7}(q,z) &= (q)_\infty^6\sum_{\substack{\ell_1,\cdots, \ell_7, \\ k_1, \cdots , k_7=0}}^\infty
{(-1)^{\sum_{i=1}^7 (k_i +\ell_i ) }q^{ {1\over 2} \sum_{i=1}^7 (k_i +\ell_i )  }  
q^{k_2(\ell_1+\ell_3) +k_4 (\ell_3+\ell_5) +k_6\ell_5 +k_7\ell_3} 
\over \prod_{i=1}^7 (q)_{k_i} (q)_{\ell_i}} \nonumber\\
&\times  \,(- z)^{-\ell_7+k_7} \, \delta_{k_4+k_7, \ell_4+\ell_7}\delta_{k_6-k_7,\ell_6-\ell_7}
\prod_{i\in \{1,2,3,5\}} \delta_{k_i,\ell_i}
\\
&=1+q+ (z+z^{-1})q^{3\over2} +3q^2+(2z^{-1} +2z) q^{5\over2} + (z^{-2} +6+z^2)q^3
+(5z^{-1} +5z) q^{7\over2} \nonumber \\
&+(2z^{-2} +12+2z^2 )q^4+(z^{-3} +10z^{-1} +10 z+z^3 )q^{9\over2}
+(6z^{-2} +21+6z^2)q^5+\cdots\,. \nonumber
\end{align}
The linear term $q$ comes from the $U(1)$ flavor symmetry current multiplet while one of the $q^2$ terms comes from the stress-energy tensor multiplet.

\subsubsection{$E_{8}$}\label{sec:E8}

\begin{figure}[h]
\begin{center}
\includegraphics[width=.7\textwidth]{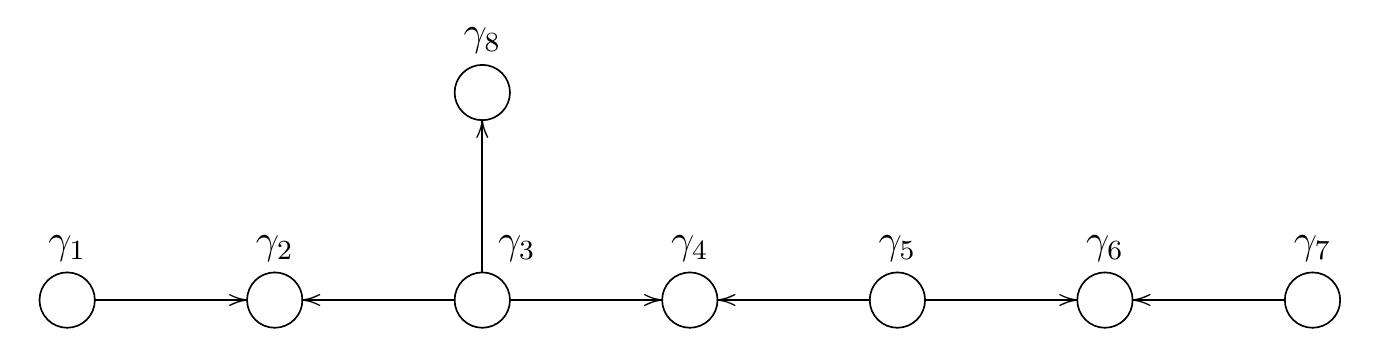}
\end{center}
\caption{ The BPS quiver for the $E_8$ Argyres-Douglas theory in the sink/source chamber.}\label{fig:E8}
\end{figure}

The $E_8$ Argyres-Douglas theory has no flavor symmetry. The 4$d$ central charge is $c_{4d}= {23\over12}$ \cite{Xie:2012hs,Xie:2013jc}. Using the 4$d$/2$d$ relation, we obtain the 2$d$ central charge,
\begin{align}
c_{2d} = - 23\,.
\end{align}
We note that $-23$ is also the central charge of the (3,8) minimal model of $W_3$ algebra \eqref{W3central}. Indeed we will show that the Schur index of the $E_8$ Argyres-Douglas theory equals to the vacuum character of the  (3,8) $W_3$ minimal model.

The Schur index of the $E_8$ Argyres-Douglas theory can be derived in a similar way as $A_{2n}$ and the $E_6$ case,
\begin{align}
\begin{split}
\mathcal{I}_{E_8}(q)&
= (q)_\infty^{8}\sum_{\ell_1,\cdots,\ell_{8}=0 }^\infty
{ q^{  \sum_{i=1}^{8} \ell_i  \,+{1\over2} \sum_{i,j=1}^{8} b^{E_8}_{ij}\ell_i \ell_{j}} \over\prod_{i=1}^{8} [(q)_{\ell_i}]^2 }\,,\\
&= 1+0q+q^2 + 2q^3 +3q^4 + 4q^5 + 7q^6 +\cdots\,,
\end{split}
\end{align}
where $b^{E_8}_{ij} =  -C^{E_8}_{ij}+2 \delta_{ij}$ and $C^{E_8}_{ij}$ is the Cartan matrix of $E_8$.

On the other hand, the vacuum character of the (3,8) $W_3$ minimal model is \cite{andrews1999a2} (see \S\ref{sec:generalAD} for more details)
\begin{align}\label{E8}
\begin{split}
\chi^{W_3,\,(3,8)} (q) &= {[(q^8;q^8)_\infty]^2\over [(q;q)_\infty]^2 } [(q;q^8)_\infty ]^2(q^6;q^8)_\infty (q^2 ;q^8)_\infty[ (q^7; q^8)_\infty]^2\\
&= 1+0q+q^2 + 2q^3 +3q^4 + 4q^5 + 7q^6 +\cdots\,,
\end{split}
\end{align}
which supports our claim
\begin{align}
\mathcal{I}_{E_8} (q) = \chi^{W_3,\,(3,8)} (q) \,.
\end{align}

\subsection{Generalized Argyres-Douglas Theories}\label{sec:generalAD}

In this subsection we will generalize our previous observations on the $A_{2n},E_6,E_8$ theories and propose that the Schur indices of the an infinite family of generalized Argyres-Douglas theories are given by the vacuum characters of $W_k$ minimal models.

The generalized Argyres-Douglas theory \cite{Cecotti:2010fi,Xie:2012hs,Xie:2013jc,DelZotto:2014kka} is defined by a BPS quiver labeled by two ADE   algebras $(G,G')$. The labeling is symmetric so the theory $(G,G')$ is the same as the theory $(G',G)$. The ordinary Argyres-Douglas theory labeled by $G$ corresponds to the special case of $(A_1,G)$.

\begin{figure}[h]
\begin{center}
\includegraphics[width=.7\textwidth]{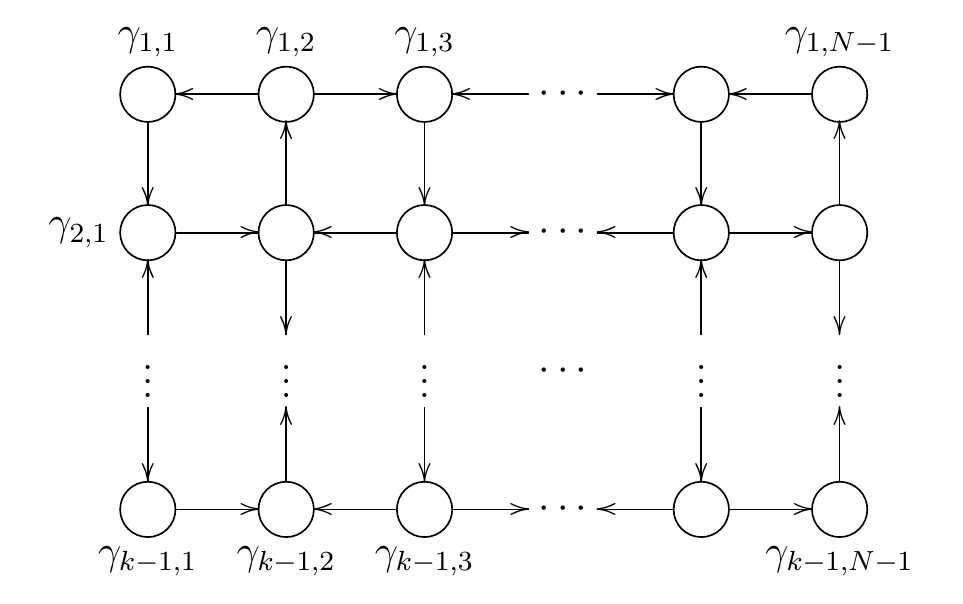}
\end{center}
\caption{ The BPS quiver for the $(A_{k-1},A_{N-1})$ Argyres-Douglas theory in the sink/source chamber.}\label{fig:AA}
\end{figure}

In particular, the BPS quiver for the  $(A_{k-1},A_{N-1})$ Argyres-Douglas theory is shown in Figure \ref{fig:AA}.   This theory has singular Seiberg-Witten curve
\begin{equation}
x^{k}+y^{N}=0~. \label{curve}
\end{equation}
The rank of the flavor symmetry is given by \cite{DelZotto:2014kka}
\begin{align}
\text{gcd}(k,N)-1\,.
\end{align}
In the following we will focus on the special case where there is \textit{no} flavor symmetry, \textit{i.e.}
\begin{align}
(A_{k-1},A_{N-1})~~~~~~~\text{with}~~\text{gcd}(k,N)=1\,.
\end{align}
The BPS states of these models are more complicated that the models studied in previous sections \cite{Cecotti:2014zga}.  In particular there are more stable BPS states than nodes of the quiver and we do not know directly how to evaluate $\mathrm{Tr}[\mathcal{O}(q)]$.  Nevertheless, we can make a reasonable conjecture as to the result.

The 4$d$ central charge of these theories are extracted from \cite{Xie:2013jc}, $c_{4d}(k,N) =  { (k-1)(N-1) (k+N+kN)\over 12 (k+N) }$ for $k$ and $N$ coprime. Using the 4$d$/2$d$ relation $c_{4d}=-12c_{2d}$, we determine the 2$d$ central charge of the underlying chiral algebra to be
\begin{align}\label{c2dAA}
c_{2d}(k,N) = -   { (k-1)(N-1) (k+N+kN)\over  (k+N) }\,,~~~~~\text{if}~~\text{gcd}(k,N)=1\,.
\end{align}
 Note that the central charge is symmetric in $k$ and $N$, $c_{2d}(k,N) = c_{2d}(N,k)$, reflecting the fact that the generalized Argyres-Douglas theory labeled by $(A_{k-1},A_{N-1})$ is the same as $(A_{N-1},A_{k-1})$.

On the other hand, the minimal model of the $W_k$ algebra is labeled by two coprime  integers $p,p'$ with $p',p\ge n$, whose central charge is given by  \cite{Fateev:1987vh,Fateev:1987zh}
\begin{align}
c^{(k)}(p,p') =  (k-1) \left[  1- k(k+1) {(p-p')^2\over p p'}\right]\,.
\end{align}
We recognize that the 2$d$ central charge \eqref{c2dAA} of the $(A_{k-1},A_{N-1})$ generalized Argyres-Dougals theory is equal to that of the $(k,k+N)$ $W_k$ minimal  model,
\begin{align}
c_{2d}(k,N)  = c^{(k)}(k,k+N)\,.
\end{align}
We therefore conjecture that the chiral algebra of the $(A_{k-1},A_{N-1})$ Argyres-Douglas theory is given by vacuum sector of  the $(k,k+N)$ $W_k$ minimal model. 

Since the $(A_{k-1},A_{N-1})$ Argyres-Douglas theory is the same as that labeled by $(A_{N-1},A_{k-1})$, for this conjecture to be true, there must be some equivalences between the $W$ algebra minimal models. Indeed, using their coset constructions, it has been shown that \cite{Altschuler:1990th},
\begin{align}
(k,k+N)~~W_k~\text{minimal model} ~= ~(N,N+k)~~W_N~\text{minimal model} \,.
\end{align}
In particular, 
\begin{align}
c^{(k)}(k,k+N) = c^{(N)}(N,N+k)\,.
\end{align}
This equivalence further solidifies our proposal.

A consequence of our conjectured identification of the Schur operators in the $(A_{N-1},A_{k-1})$ Argyres-Douglas theory as the vacuum sector of the $(k,k+N)$ $W_{k}$ minimal model is that Schur index must equal to the associated vacuum character which has a compact closed form expression. In general the character, or equivalently, the highest weight of the $(k,k+N)$ $W_k$ minimal model is labeled by a tuple of positive integers 
\begin{align}
(j_0,j_1,\cdots, j_{k-1})~,
\end{align}
 such that $\sum_{i=0}^{k-1}j_i=k+N$. The labeling is not unique and  we need to identify the two highest weights $(j_0,j_1,\cdots,j_{k-1})$ if they differ by a $\mathbb{Z}_k$ cyclic permutation \cite{Bouwknegt:1992wg}.

Explicitly, the character of the highest weight module labeled by $(j_0,j_1,\cdots, j_{k-1})$ in the  $(k,k+N)$ $W_k$ minimal model is given by \cite{andrews1999a2} (see also \cite{LF,Bouwknegt:1991gf,frenkel1992characters,Bouwknegt:1992wg}),
\begin{align}\label{Wcharacter}
\chi ^{W_k,\,(k,k+N)}_ {(j_0,j_1,\cdots,j_{k-1})} (q)
= \left[{ (q^{k+N} ; q^{k+N} )_\infty\over (q)_\infty}\right]^{k-1}
\prod_{a=1}^{k-1}\prod_{b=0}^{k-1} (q^{j_b+j_{b+1}+\cdots +j_{a+b-1}};q^{k+N})_\infty\,,
\end{align}
where we define $j_i=j_{i'}$ if $i\equiv i'$ (mod $n$). Indeed, the character is manifestly cyclic invariant in $j_i$'s.

The $(k,k+N)$ $W_k$ minimal model  character that is of particular interest to us is the vacuum character, which is labeled by
\begin{align}
j_0 = N+1,~j_1=j_2=\cdots=j_{k-1}=1\,,
\end{align}
or a cyclic permutation thereof. One can check that the dimension of the highest weight state is indeed zero \cite{Bouwknegt:1992wg}. Then, from \eqref{Wcharacter}, the vacuum character  is
\begin{align}
\begin{split}
\chi^{W_k,\, (k,k+N)}_0(q)&\equiv\chi ^{W_k,\,(k,k+N)}_ {(N+1,1,\cdots,1)} (q)\\
&=\left[{ (q^{k+N} ; q^{k+N} )_\infty\over (q)_\infty}\right]^{k-1}
\prod_{a=1}^{k-1}\, [ (q^{N+a};q^{k+N})_\infty ]^a[(q^{a};q^{k+N})_\infty ]^{k-a}\,.
\end{split}
\end{align}
We therefore have the following conjecture for the Schur index of the $(A_{k-1},A_{N-1})$ generalized Argyres-Douglas theory,
\begin{align}\label{Wconjecture}
\textbf{Conjecture}:~\mathcal{I}_{(A_{k-1},A_{N-1})}(q)  =  \chi^{W_k,\, (k,k+N)}_0(q)\,,~~~~\text{if}~~\text{gcd}(k,N)=1\,.
\end{align}

Our previous calculations may be interpreted as evidence for this proposal.  Indeed:
\begin{itemize}
\item In \S\ref{sec:A2n} we observed that 
\begin{align}
\mathcal{I}_{A_{N-1} \cong (A_1,A_{N-1})}(q) = \chi^{W_2,\, (2,N+2)}_0(q)\,,~~~~\text{for}~~N~\text{odd}\,.
\end{align}
\item Next, consider the $(A_2,A_3)$ theory. In this case the singularity \eqref{curve} is equivalent to the $E_{6}$ singularity.  Moreover the equivalences between the two BPS quivers may be demonstrated by mutation. In \S\ref{sec:E6}, we found 
\begin{align}
\mathcal{I}_{E_6\cong (A_2,A_3)}(q) = \chi^{W_3,\, (3, 7)}_0(q)\,.
\end{align}
\item Finally, consider the $(A_2,A_4)$ theory. Again in this case we recognize the singularity to be of $E_{8}$ type and again the two quivers are mutation equivalent. In \S\ref{sec:E8}, we found 
\begin{align}
\mathcal{I}_{E_8\cong (A_2,A_4)}(q) = \chi^{W_3,\, (3, 8)}_0(q)\,.
\end{align}
\end{itemize}
It would be interesting to evaluate the trace of $\mathcal{O}(q)$ and verify this idea.

\section*{Acknowledgements} 
We thank Chris Beem, Nathan Benjamin, Chi-Ming Chang, Thomas Dumitrescu, Daniel Jafferis, Ying-Hsuan Lin, and Andrew Neitzke,  Natalie M. Paquette, Cumrun Vafa, Dan Xie, and Xi Yin for discussions.  The work of CC is support by a Junior Fellowship at the Harvard Society of Fellows.  The work of SHS is supported by the Kao Fellowship at Harvard University.

\bibliography{WCindex}{}
\bibliographystyle{utphys}

\end{document}